\documentclass[usenatbib, usegraphicx]{mn2e}
\usepackage{amssymb}
\usepackage{amsmath}
\usepackage{array}
\usepackage{bigstrut}
\usepackage{cellspace}
\usepackage{graphicx}
\newif\ifAMStwofonts
\def\apj{ApJ}
\def\mnras{MNRAS}

\def\aap{A\&A}

\def\ut #1 #2 { \, \rmn{#1}^{#2}}
\def\u #1 { \, \rmn{#1}}

\def\grad{\bmath{\nabla}}
\def\cross{\bmath{\times}}
\def\bdot{\bmath{\cdot}}
\def\curl #1 {\grad \cross #1}
\def\div #1 {\grad \bdot #1}

\def\b{\bmath{b}}
\def\e{\bmath{e}}
\def\v{\bmath{v}}
\def\w{\bmath{w}}

\def\B{\bmath{B}}
\def\E{\bmath{E}}            % E
\def\e{\bmath{e}}            % e
\def\we{\bmath{w}_{\rm E}}   %w_E
\def\fh{\bmath{\hat{\phi}}}  % phi
\def\rh{\bmath{\hat{r}}}     % r
\def\rhot{\tilde \rho}     % \rho tilde
\def\zt{\tilde z}          % z tilde
\def\beti{\beta_{\rm i}}  %beta_i
\def\betiabs{|\beta_{\rm i0}|}
\def\betio{\beta_{\rm i0}}
\def\bete{\beta_{\rm e}}  %beta_e
\def\beteo{\beta_{\rm e0}}
\def\lao{\Lambda_{\rm 0}}
\def\upo{\Upsilon_{\rm 0}}
\def\xib{\xi_{\rm b}}        %xi_b
\def\fb{f_{\rm b}}            % f_b
\def\vr{v_{r}}               % v_r
\def\vf{v_{\phi}}            % v_phi
\def\vk{v_{\rm K}}           % v_K
\def\vz{v_{z}}               % v_z

\def\wr{w_{r}}               % w_r
\def\wf{w_{\phi}}            % w_phi
\def\wz{w_{z}}               % w_z

\def\br{b_{r}}
\def\ba{b_{\phi}}

\def\er{e_{r}^{\prime}}
\def\ef{e_{\phi}^{\prime}}
\def\ez{e_{z}^{\prime}}

\def\jr{j_{r}}
\def\jf{j_{\phi}}

\def\sigpar{\tilde{\sigma}_{\rm O}}
\def\sigP{\tilde{\sigma}_{\rm P}}
\def\sigH{\tilde{\sigma}_{\rm H}}

\def\J{\bmath{J}}
\def\j{\bmath{j}}

\newcommand{\ee}[1]{\times 10^{#1}}

\usepackage[]{graphics}

\title{Wind-driving protostellar accretion discs. II. Numerical method and illustrative solutions}
\author[R. Salmeron, A. K\"onigl and M. Wardle]
       {Raquel Salmeron$ ^{1,2} $,  Arieh K\"onigl$ ^2 $ \& Mark Wardle$ ^3 $ \\
 $ ^1 $Research School of Astronomy \& Astrophysics and Research School of Earth Sciences, The Australian National University, \\
 Canberra ACT 0200, Australia \\
 $ ^2 $Department of Astronomy \& Astrophysics and The Enrico Fermi
 Institute, The University of Chicago, Chicago  IL 60637, USA \\
$ ^3 $Department of Physics and Astronomy, Macquarie University, Sydney NSW 2109, Australia}

\pagerange{\pageref{firstpage}--\pageref{lastpage}}
\pubyear{2010}
\begin{document}

\maketitle
\label{firstpage}
\begin{abstract}
We continue our study of weakly ionized protostellar accretion discs
that are threaded by a large-scale magnetic field and power a
centrifugally driven wind. It has been argued that there is already
evidence in several protostellar systems that such a wind transports a
significant fraction of the angular momentum from at least some part of
the disc. We model this situation by considering a radially localized
disc model in which the matter is everywhere well coupled to the field
and the wind is the main repository of excess angular momentum. We
consider stationary configurations in which magnetic diffusivity
counters the shearing and advection of the magnetic field lines. In
Wardle \& K\"onigl we analysed the disc structure in the hydrostatic
approximation (vertical motions neglected inside the disc) and presented
exact disc/wind solutions for the ambipolar diffusivity regime. In
K\"onigl, Salmeron \& Wardle (Paper I) we generalized the hydrostatic
analysis to the Hall and Ohm diffusivity domains and used it to identify
the disc parameter sub-regimes in which viable solutions with distinct
physical properties can be expected to occur. In this paper we test the
results of Paper I by deriving full numerical solutions (integrated
through the sonic critical surface) of the disc equations in the Hall
domain. We confirm all the predictions of the hydrostatic analysis and
demonstrate its usefulness for clarifying the behaviour of the derived
solutions. We further show that the outflow solutions can be extended to
larger scales (so that, in particular, they also cross the Alfv\'en
critical surface) by matching the localized disc solutions to global
`cold' wind solutions of the type introduced by Blandford \& Payne. To
facilitate this matching, we construct a library of wind solutions for a
wide range of wind model parameters; this library is made available to
the community.

The results presented in Wardle \& K\"onigl, Paper I and this work
combine to form a comprehensive framework for the study of wind-driving
accretion discs in protostellar and other astrophysical
environments. This theoretical tool could be useful for interpreting
observations and for guiding numerical simulations of such systems.
\end{abstract}

\section{Introduction}
\label{sec:intro}

A common feature of protostellar accretion discs is their association
with collimated, energetic outflows \citep[e.g.][]{C07}. These
outflows are widely believed to represent centrifugally driven winds
that are launched along large-scale, ordered magnetic fields
\citep[e.g.][]{KP00,POFB07}. Such winds could in principle be efficient
trasnporters of disc angular momentum \citep[e.g.][hereafter
BP82]{BP82}, and it has in fact been argued that there is already
observational evidence in several protostellar systems that an outflow
of this type carries the bulk of the excess angular momentum from at
least some part of the associated accretion disc
\citep[e.g.][]{Ray07}. Protostellar discs are typically weakly ionized,
and a certain minimum degree of ionization is required to attain the
level of coupling between the matter and the field that enables the
vertical magnetic angular-momentum transport mechanism to operate. [A
similar requirement must be satisfied also to enable radial
angular-momentum transport by a small-scale, turbulent magnetic field;
such turbulence could be induced, for example, by the magnetorotational
instability (MRI; e.g. \citealt{BH98}).] The inherent magnetic
diffusivity tends to counter the effects of shearing and advection by
the differentially rotating accretion flow and therefore makes it
possible for the discs to attain a steady state, at least on the
dynamical (rotation) time.

The behaviour of a weakly ionized gas can be characterized according to
the nature of the dominant diffusivity mechanism for the given density
and ionization state: ambipolar, Hall, or Ohm \citep[e.g.][]{KS11}. The
ambipolar regime, in which the magnetic field lines are effectively
frozen into the ions and drift with them relative to the dominant
neutral component, can be expected to dominate over the entire disc
cross section in the outermost regions of the disc (at radii $r \gtrsim
10\;$au) and near the disc surfaces at smaller radii. In the Hall regime
the magnetic field is frozen into the electrons and drifts with them
relative to the ions and neutrals; this regime could dominate over most
of the disc cross sections on scales $r\sim 1-10\;$au. The Ohm regime,
in which the field lines completely decouple from the charge carriers,
could potentially dominate near the disc mid-plane on scales $\sim
0.1-1\;$au. (At smaller radii the mid-plane region likely becomes
collisionally ionized.) However, as wind-driving discs typically have
comparatively lower column densities and mid-plane densities because
of their high angular momentum transport efficiency (which results in
relatively high inflow speeds), the low-ionization Ohm regime might be
curtailed (or even entirely eliminated) in such systems. Even if this
were to happen, the Hall-dominated zone would be unlikely to extend to
much smaller radii than estimated above on account of the fact that, at
sufficiently large densities, the dominant charge carriers become
positively and negatively charged grains of equal mass
\citep*[e.g.][]{NNU91}, so that no Hall current can flow.

The structure of radially localized wind-driving protostellar discs in
the ambipolar diffusion-dominated regime was investigated by
\citet[][hereafter WK93]{WK93}. They derived exact solutions that were
matched to global, radially self-similar disc-wind solutions of the type
introduced by BP82. They analysed the disc solutions by using the
hydrostatic approximation, in which the vertical velocity component is
neglected inside the disc, and obtained useful algebraic relations that
led to a set of parameter constraints on physically viable
configurations \citep[see also][]{Kon97}. This analysis was generalized
by \citet*[][hereafter Paper I]{KSW10} to the Hall and Ohm diffusivity
regimes. In particular, they found that all the viable solutions
correspond to four sub-regimes in the Hall domain and three in the Ohm
domain, with the solutions in each of the identified parameter sectors
having distinct physical properties. The main goal of this paper is to
derive exact wind-driving disc solutions that can test these predictions
and, more generally, the applicability of the hydrostatic approximation
to the description of the salient features of such systems. We
concentrate on the Hall diffusivity domain, which, according to the
discussion above, should be most relevant (together with the ambipolar
regime already discussed in WK93) to the study of the weakly ionized
regions of wind-driving protostellar discs. In Paper I we formulated the
problem in terms of a conductivity tensor, specified by the values of
the Pedersen, Hall and Ohm components, and discussed its correspondence
to a multifluid formulation for the case where only two charged species
(one positive and one negative) are present. Our parameter constraints were, in
fact, derived in the context of the latter approach. In this paper
we employ the conductivity-tensor formulation for the characterization
of the solutions that we derive, but we again employ the multifluid
formulation in the analysis.

The paper is organized as follows. Section~\ref{sec:local} summarizes
the system of equations that underlies the radially localized disc model
and describes the model parameters as well as the boundary conditions on
the equations and the method of their numerical
integration. Section~\ref{sec:wind} provides the corresponding
description of the global, self-similar wind model and outlines the
procedure used to match the disc and wind
solutions. Section~\ref{sec:results} presents representative solutions
for Hall diffusivity-dominated discs in the different parameter-space
sub-regimes identified by the hydrostatic analysis in Paper~I and
compares them with the predictions (summarized in
Appendix~\ref{sec:appA}) of the analytical treatment. The main
findings of the paper are recapitulated in Section~\ref{sec:conclude}.

\section{Radially localized disc models}
	\label{sec:local}
\subsection[]{Dimensionless system of equations in $z$}
	\label{subsec:gov}

In Section~I.3 (where the prefix `I' hereafter denotes Paper I) we
derived the algebraic relations and ordinary differential equations
(ODEs) in the vertical cylindrical coordinate $z$ that describe the
vertical structure of the disc at any given value of the radial
cylindrical radius $r$. For clarity and ease of reference, we reproduce
them below in dimensionless form. They are the equations of motion
\begin{equation} 
\frac{d w_r}{d\zt} = \frac{1}{\wz}\left[\frac{a_0^2}{\rhot} \jf + 2
\wf\right]\, ,
\label{eq:r_motiond} 
\end{equation} 
\begin{equation} 
\frac{d w_{\phi}}{d\zt} = -\frac{1}{w_z} \left[\frac{a_0^2}{\rhot} \jr +
\frac{w_r}{2}\right]\, ,
\label{eq:phi_motiond} 
\end{equation}
\begin{equation} 
\frac{d \ln{\rhot}}{d\zt} = \frac{1}{1 - w_z^2}
\left[\frac{a_0^2}{\rhot} (\jr \ba - \jf \br) - \zt \right] \,;
\label{eq:z_motiond} 
\end{equation}
the azimuthal component of the induction equation 
\begin{equation} \frac{d w_{{\rm E}r}}{d\zt} =
-\frac{3}{2} \br \, ; 
\label{eq:phi_inductiond} 
\end{equation} 
Amp\`ere's Law 
\begin{equation} 
\frac{d \br}{d\zt} = \jf\, ,
\label{eq:phi_ampered}
\end{equation} 
\begin{equation} 
\frac{d \ba}{d\zt} = - \jr \,;
\label{eq:r_ampered} 
\end{equation} 
the relations linking the electric field in the inertial coordinate
system and in the frame comoving with the neutrals
\begin{equation} 
\er = w_{{\rm E}r} + \wf - \wz\ba\, , 
\label{eq:E-E'} 
\end{equation} 
\begin{equation} 
\ef = -\epsilon_{\rm B} + w_z \br - w_r \,; 
\label{eq:E-E'1} 
\end{equation} 
and Ohm's Law
\begin{equation} \jr = y (\sigpar - \sigP) \br + \frac{\sigH}{b} (\ez
\ba - \ef) + \sigP \er\, ,
\label{eq:jr} 
\end{equation} 
\begin{equation} 
\jf = y (\sigpar - \sigP) \ba + \frac{\sigH}{b} (\er - \ez \br) + \sigP
\ef\, ,
\label{eq:jphi}
\end{equation}
\begin{eqnarray} 
\lefteqn{\ez = \frac{-(\er \br + \ef \ba)(\sigpar - \sigP)}{ (\sigpar -
\sigP) + b^2 \sigP} + {} } \nonumber\\ 
& & {} \frac{\sigH b (\er \ba - \ef \br)}{ (\sigpar - \sigP)
+ b^2 \sigP} \, ,  
\label{eq1:ez} 
\end{eqnarray} 
where we have taken $j_z$ to be $\simeq 0$.
The following notation has been used in the above expressions:
\begin{equation}
\label{eq:nondim1} 
\zt \equiv \frac{z}{h_{\rm T}}\ , \qquad \rhot \equiv
\frac{\rho(r,z)}{\rho_0(r)}\ , 
\end{equation} 
\begin{equation}
\label{eq:nondim2} 
\w \equiv \frac{\v - \vk \fh}{c_{\rm s}}\ , \ \we
\equiv \frac{c \E/B_0 + \vk \rh}{c_{\rm s}}\ , \ \e^\prime \equiv \frac{c
\E^\prime}{c_{\rm s} B_0} \ , 
\end{equation} 
\begin{equation}
\label{eq:nondim3} \j \equiv \frac{4 \pi h_{\rm T} \J}{c B_0}\ , \quad
\bmath{\tilde \sigma} \equiv \frac{4 \pi h_{\rm T} c_{\rm
s}\bmath{\sigma}}{c^2}\ , \quad \b \equiv \frac{\B}{B_0} \ , 
\end{equation}
where $\rho$ is the gas density, $\bmath{v}$ is the fluid velocity,
$\bmath{J}$ is the current density, $\bmath{B}$ is the magnetic field,
$\bmath{\sigma}$ is the conductivity tensor, incorporating the Pedersen,
Hall and Ohm components ($\sigma_{\rm P}$, $\sigma_{\rm H}$ and
$\sigma_{\rm O}$, respectively), and 
\begin{equation} \bmath{E'} = \bmath{E} + \frac{\bmath{v} \times
\bmath{B}}{c} 
\end{equation} 
is the electric field in the frame of the neutrals, in terms of $\bmath{E}$,
the electric field in the inertial (`laboratory') frame. Also, $y\equiv
\E^\prime \cdot \B/B^2$, $v_{\rm K}$ is the Keplerian speed, $c_{\rm s}$
is the isothermal speed of sound, $h_{\rm T}=(c_{\rm s}/v_{\rm K})r$ is
the tidal scaleheight and the subscript `0' denotes the disc mid-plane.

Under the thin-disc approximation, the vertical component of the
magnetic field is constant with height; there is therefore no need to
solve the vertical component of Amp\'ere's Law. It can also be shown
that the azimuthal component of $\E$ (or, equivalently, the variable
$w_{{\rm E}r}$) is constant with height in our model (see
Section~I.3.7). Therefore the variable $v_{Br} = cE_\phi/B_Z$, which
represents the radial drift velocity of the poloidal flux surfaces (see
Section~I.3.6) is also constant with height inside the disc.
 
In addition to the above equations, the following are also included so
as to enable the position of the sonic point (subscript `s') and the
(normalized) upward mass flux $w_{z{\rm 0}}$ ($= \rhot_{\rm s}$ under
the assumptions that $\rhot w_z = {\rm const}$ and that the disc is
vertically isothermal) to be derived self-consistently as part of
the solution (see Section~\ref{subsec:discsol}):
\begin{equation} 
\frac{d \zt_{\rm s}}{d\zt} = 0 \, ,
\label{eq:z_s} 
\end{equation}
\begin{equation} 
\frac{d w_{z{\rm 0}}}{d\zt} = 0 \,.
\label{eq:w0} 
\end{equation}

\subsection{Parameters}
	\label{subsec:param}

The solutions are specified by the parameters enumerated below. Only a
brief summary is provided here; the reader is referred to Section~I.3.13
for further details.

\begin{enumerate}

\item $a_{\rm 0} \equiv v_{{\rm A}z0}/c_{\rm s}$, the mid-plane ratio of
the Alfv\'en speed (based on the uniform vertical field component $B_z$)
to the isothermal sound speed. It is a measure of the magnetic field strength.

\item $c_{\rm s}/\vk = h_{\rm T}/r$, the ratio of the tidal scaleheight
to the disc radius -- a measure of the geometric thinness of the
disc. Although this parameter does not appear explicitly in the
normalized equations, it is 
required for mathcing the disc and wind solutions (see
Sections~\ref{subsec:wind_param} and~\ref{subsec:globalsol}) and is used
to place an upper limit on the midplane radial speed in physically
viable solutions (see Appendix~\ref{sec:AppA}).

\item The mid-plane ratios of the conductivity tensor components: $(\sigma_{\rm
P}/\sigma_\perp)_0$ (or $(\sigma_{\rm H}/\sigma_\perp)_0$) and
$(\sigma_\perp/\sigma_{\rm O})_0$, where $\sigma_\perp = (\sigma_{\rm H}^2 +
\sigma_{\rm P}^2)^{1/2}$ is the total conductivity perpendicular to the
magnetic field. They characterize the conductivity regime of the fluid
and are taken here to be independent of $z$ to facilitate the comparison
with the analytic results derived in Paper~I.

\item The mid-plane value of the Elsasser number $\Lambda_{\rm 0} \equiv
v_{{\rm A}0}^{2}/(\Omega_{\rm K}\eta_{\perp 0})$, where $\eta_{\perp{\rm
0}} \equiv c^2/4\pi \sigma_{\perp {\rm 0}}$ is the `perpendicular'
magnetic diffusivity and $\Omega_{\rm K}=v_{\rm K}/r$ is the Keplerian
angular velocity. This parameter measures the degree of coupling between
the neutrals and the magnetic field, with values $\gg 1$ and $\ll 1$
corresponding to strong and weak coupling, respectively.  In this paper
we assume (as was done in paper~I) that the disc is everywhere
magnetically well coupled, so that $\Lambda _0$ is never $\ll 1$.  As an
alternative to the tensor magnetic diffusivity employed here, one could
also write down separate equations for the charged species (see
Paper~I). Assuming that the charged particles consist only of ions and
electrons (denoted by the subscripts `i' and `e', respectively), one can
write $\Lambda = \Upsilon \beta_{\rm i}$ in the Hall regime, where
$\Upsilon$ is the ratio of the Keplerian rotation time to the
neutral-ion momentum exchange time and $\beta_j$ is the ratio of the
gyrofrequency of charged species $j$ to its collision frequency with the
neutral gas (see Appendix~\ref{sec:appA}). In the ambipolar diffusivity
limit (not considered here), $\Lambda = \Upsilon$.

\item $\epsilon \equiv -v_{r0}/c_{\rm s}$, the normalized inward radial
speed at the mid-plane. This is a free parameter of the \emph{disc
solution}, whose value is determined when it is matched to a self-similar
global wind solution (see Section~\ref{subsec:globalsol}).

\item $\epsilon_{\rm B} \equiv -v_{Br0}/c_{\rm s} = -cE_{\phi 0}/c_{\rm
s} B_z$, the normalized (vertically constant) azimuthal component of the
electric field, which measures the radial drift velocity of the poloidal
magnetic field lines through the disc.

\end{enumerate}   
  
\subsection{Boundary Conditions}
	\label{subsec:boundary}

The complete system of equations comprises the set of ODEs given by
equations~(\ref{eq:r_motiond})--(\ref{eq:r_ampered}), (\ref{eq:z_s})
and~(\ref{eq:w0}) as well as the algebraic relations
(\ref{eq:E-E'})--(\ref{eq1:ez}). This is a two-point boundary value
problem for coupled ODEs. Eight boundary conditions must be formulated,
either at the mid-plane or at the critical (sonic) point of the flow,
defined as the location where the vertical velocity reaches the
isothermal sound speed (or $w_z = 1$). They are applied as follows.
\begin{description} 
\item \textbf{At the mid-plane}. We begin by
assigning odd symmetry to the radial and azimuthal components of the
magnetic field, $ \br$ and $\ba$, which therefore vanish at $\tilde
z=0$. Consistently with this choice, the remaining variables (except
$w_z$) have even (reflection) symmetry about the mid-plane and their
derivatives vanish at that location.  We also adopt $\tilde \rho=1$,
which follows directly from the normalization of $\rho$, and prescribe the
radial inward velocity (measured by the parameter $\epsilon$). The six
boundary conditions applied at the mid-plane are thus
\begin{equation}
b_{r{\rm 0}} = b_{\phi{\rm 0}} = 0 \, ,
\label{eq:b1}
\end{equation}
\begin{equation}
\left (\frac{d w_r}{d \tilde z}\right )_{\rm 0} = \left
(\frac{dw_{\phi}}{d \tilde z}\right )_{\rm 0} = 0 \, ,
\label{eq:b2}
\end{equation}
\begin{equation}
\tilde \rho_{\rm 0} = 1 \, , \qquad -w_{r{\rm 0}} = \epsilon \, .
\label{eq:b3}
\end{equation}

Using these boundary conditions and equations~(\ref{eq:r_motiond}),
(\ref{eq:phi_motiond}) and~(\ref{eq:E-E'})--(\ref{eq:jphi}), we arrive
at the following expressions for $w_{\phi{\rm 0}}$ and $w_{{\rm E}r{\rm
0}}$:
\begin{equation}
w_{\phi{\rm 0}} = - \frac{\sigma_{\rm H0}}{\sigma_{\rm
P0}}\frac{\epsilon}{4} - \frac{a_0^2 \tilde \sigma_{\perp{\rm 0}}^2}{2 
\sigma_{\rm P0}} (\epsilon - \epsilon_{\rm B})\, ,
\label{wphi0} 
\end{equation}
\begin{equation}
w_{{\rm E}r{\rm 0}} = \frac{\sigma_{\rm H0}}{\sigma_{\rm P0}} (\epsilon
- \epsilon_{\rm B}) + \frac{1}{\sigma_{\rm P0}}\frac{\epsilon}{2
a_0^2} - w_{\phi{\rm 0}} \, . 
\label{eq:b4}
\end{equation}
\item \textbf{At the sonic point}: The sonic point is a singular
point of equation~(\ref{eq:z_motiond}), so an additional boundary
condition is obtained by imposing the regularity condition
\begin{equation}
\tilde z_{\rm s} = \frac{a_0^2}{\tilde \rho_{\rm s}}  (j_{r{\rm s}}
b_{\phi{\rm s}} - j_{\phi{\rm s}} b_{r{\rm s}}) \,.
\label{eq:sonic}
\end{equation}
Furthermore, the density at $\tilde z_{\rm s}$ is given by
\begin{equation}
\tilde \rho_{\rm s} = w_{z{\rm 0}}
\label{eq:rs}
\end{equation}
(see the discussion just before equation \ref{eq:z_s}). The density
derivative at the sonic point can be obtained by differentiating the
numerator and denominator of equation~(\ref{eq:z_motiond}) (i.e. by
applying l'H\^{o}pital's rule), which yields a quadratic equation for
$(d\rhot/d \tilde z)_{\rm s}$. Out of the two roots of this equation, we
choose the negative one -- corresponding to a positive velocity gradient
-- as expected for a physical solution.
\end{description}

\subsection{Numerical integration of the disc equations}
\label{subsec:discsol}

The set of ODEs given by
equations~(\ref{eq:r_motiond})--(\ref{eq:r_ampered}), (\ref{eq:z_s})
and~(\ref{eq:w0}) is integrated using the procedure first outlined in
WK93. In brief, we start by assigning the mid-plane values of $b_r$,
$b_{\phi}$, $\rhot$, $w_r$, $w_{\phi}$ and $w_{{\rm E}r}$ (using
equations~\ref{eq:b1} and \ref{eq:b3}--\ref{eq:b4}), proceed to guess
the value of $w_{z{\rm 0}}$ (or, equivalently, $\tilde \rho_s$) as well
as the position of the sonic point $\tilde z_{\rm s}$, and then
integrate from the mid-plane vertically upward. If the guessed $w_{z{\rm
0}}$ is too high, $w_z$ eventually diverges. If, however, the guessed
value is too low, the vertical velocity reaches a maximum and then
begins to decrease with $\zt$. But a vertically decreasing velocity
corresponds to a positive density gradient, which is unphysical for an
isothermal disc in which the nonthermal forces acting in the vertical
direction (namely, the vertical components of gravity and of the Lorentz
force) tend to compress the gas. This behaviour of the solutions enables
us to bracket the correct value of $w_z$ between two limits. We then
improve on the guessed value using bisection until we are close enough
to the physical solution to be able to estimate $\tilde z_s$ and hence
(by extrapolating in $\zt$) the values of the fluid variables at the
sonic point. The boundary conditions given by equations~(\ref{eq:sonic})
and~(\ref{eq:rs}) can then be imposed to obtain the remainder of the
disc variables at that location.

To complete the derivation of the full solution, we renormalize $z$ by
the estimated height of the sonic point ($\hat{z} \equiv z/z_{\rm s}$)
and integrate the equations simultaneously from the mid-plane ($\hat{z}
= 0$) and from the sonic point ($\hat{z}_{\rm s} = 1$) to an
intermediate fitting point (typically $\hat{z} \sim 0.90$) while
adjusting the guessed values at both locations iteratively until the
solution converges. This procedure is adopted because the integration
from the sonic point toward the mid-plane becomes unstable at small
values of $\hat{z}$.  In carrying out this numerical scheme, it is
essential to ensure that $w_z$ attains a value of at least $\sim 0.90 -
0.95$ in the bisection runs before attempting to compute the location of
the sonic point and the extrapolated values of the fluid variables
there. If this is not done then the estimates of these quantities may
not be good enough for the backward integration to the fitting point to
be successful.

\section{Global (self-similar) wind models}
\label{sec:wind}

The methodology described in Section~\ref{sec:local} is appropriate for
determining the initial acceleration of the wind from the surfaces of
the disc, but it cannot be used to follow the evolution of the outflow
on scales where the adopted thin-disc approximation
$z/r \ll 1$ breaks down. It is, however, necessary to ensure
that the obtained wind solution continues to accelerate and passes
through the remaining critical surfaces of the problem, which correspond
to the Alfv\'en and fast-magnetosonic critical points
\citep[e.g. BP82;][]{VTST00,FC04}.
Since we cannot self-consistently model this process in view of our
radially localized formulation of the disc structure, we approximate a
global disc--wind solution by matching the disc solution to a global,
radially self-similar (i.e. BP82-type) wind solution. In our cold-wind
approximation we only need to impose the regularity condition at the
Alfv\'en critical surface, which allows us to constrain one of the disc
parameters (specifically $\epsilon$, which quantifies the mid-plane
radial velocity). We further simplify the treatment by assuming that
$\epsilon_{\rm B} = 0$, which allows us to avoid relating the field-line
inclination at the disc surface to the global magnetic flux distribution
along the disk \citep[see][]{OL01}.  This is essentially the procedure
employed (in the ambipolar diffusivity regime) by WK93. In
sections~\ref{subsubsec:selfsimilar}--\ref{subsec:wind_int} we describe,
in turn, the governing equations, parameters, boundary conditions and
numerical integration of the self-similar wind solution. The methodology
for matching a localized disc solution to a global wind solution is
presented in Section~\ref{subsec:globalsol}.

\subsection{Governing equations}
\label{subsubsec:selfsimilar}

The wind is described by the steady-state, axisymmetric, ideal-MHD
equations \citep[e.g.][]{S93,KS10} that comprise the conservation of mass
\begin{equation}
\label{eq:wind1}
\nabla \cdot ( \rho \bmath{v}) = 0
\end{equation}
and momentum
\begin{equation}
\label{eq:wind2}
\rho \bmath{v} \cdot \nabla \bmath{v} = - \nabla P - \rho \nabla \Phi
+ \frac{ \bmath{J} \times \bmath{B}}{c}
\end{equation}
for the neutral gas, the induction equation for the topology of the
magnetic field
\begin{equation}
\label{eq:wind4}
\nabla \times (\bmath{v} \times \bmath{B} ) = 0 \; ,
\end{equation}
Amp\`ere's Law
\begin{equation}
\label{eq:ampere}
\bmath{J} = \frac{c}{4 \pi}  \nabla \times \bmath{B}
\end{equation}
(where, as customary, we have neglected the displacement current), and
the solenoidal condition on the magnetic field ($\nabla \cdot \bmath{B}
= 0$). In the above expressions, $P$ is the gas pressure and $\Phi$ is
the gravitational potential of the central object,
\begin{equation}
\label{eq:phi}
 \Phi = - \frac{GM}{(r^2 + z^2)^{1/2}}    \;,
 \end{equation}
where $G$ is the gravitational constant and $M$ is the mass of the
protostar. In ideal-MHD flows, the magnetic field and velocity vectors
are parallel in a frame that moves with the angular velocity
$\bmath{\Omega}_{\rm B}$ 
of the magnetic flux surfaces,
\begin{equation}
\label{eq:vB}
\bmath{v} = \frac{k \bmath{B}}{4 \pi \rho} + (\bmath{\Omega}_{\rm B}
\times \bmath{r}) \,. 
\end{equation}
where $k/4\pi$ is the \emph{mass load function} of the wind (the
ratio of the constant mass flux to the constant magnetic flux). In the
poloidal ($r-z$) plane (subscript `p'), this equation reduces to
\begin{equation}
\bmath{v}_{\rm p} = \frac{k \bmath{B}_{\rm p}}{4 \pi \rho} \, .
\end{equation}
The variables $\Omega_{\rm B}$ and $k$ satisfy $(\bmath{B} \cdot \nabla)
\Omega_{\rm B} = (\bmath{B} \cdot \nabla) k = 0$, and thus are constant
along the magnetic field lines (or, equivalently, the wind
flowlines). Additional quantities that remain constant along the flow
are the specific energy
\begin{equation}
\label{eq:energy}
e =  \frac{1}{2} v^2 + h + \Phi  -\frac{\Omega_{\rm B} r B_\phi}{k} \,,
\end{equation}
where $h$ is the enthalpy per unit mass, and the specific angular momentum
\begin{equation}
\label{eq:wind9}
l = r v_{\phi} - \frac{r B_{\phi}}{k} \,,
\end{equation}
which incorporates the contributions of both the matter (the first term on the
right-hand side) and the magnetic field (the second term). The
quantities $k$, $e$ and $l$ can be expressed in dimensionless form as 
\begin{equation}
\kappa \equiv k \ (1 + \xi'^2_{\rm b})^{1/2} \ 
\frac{v_{\rm Kb} }{B_{\rm b}}\, ,
\end{equation}
\begin{equation}
\label{eq:energy2}
\varepsilon \equiv \frac{e}{v_{\rm Kb}^2} 
\end{equation}
and
\begin{equation}
\label{eq:l2}
\lambda \equiv \frac{l}{\sqrt{GMr_{\rm b}}} \; ,
\end{equation}
where the subscript `b' denotes the location of the base of the wind.

We now introduce the self-similarity scalings using the notation of BP82:
\begin{equation}
r = r_{\rm b} \xi(\chi)\, ,
\label{eq:scaling1}
\end{equation}
\begin{equation}
 z =  r_{\rm b} \chi \, ,
 \end{equation}
\begin{equation}
v_r = \xi'(\chi) f(\chi) v_{\rm Kb}\, ,  
\label{eq:scalingvr}
\end{equation}
\begin{equation}
v_\phi = g(\chi) v_{\rm Kb}\, ,
\end{equation}
\begin{equation}
\label{eq:scalingvz}
v_z = f(\chi) v_{\rm Kb} \, .
\end{equation}
In these expressions, $\xi' \equiv \tan \varphi = B_r/B_z$ is the
inclination of the field lines with respect to the rotation axis of the
star and disc. At the base of the wind, we take $\chi_{\rm b} = 0$,
$\xi_{\rm b} = 1$, $g_{\rm b} = 1$ and $f_{\rm b} = 0$, so the fluid
velocity at the launching point of the outflow is exactly Keplerian.

We now sketch the procedure followed by BP82 to obtain the set of
ODEs in $\chi$ that describe the self-similar
wind solution. First, from the scalings
(\ref{eq:scalingvr})--(\ref{eq:scalingvz}), we deduce
\begin{equation}
\label{eq:vsquared}
v^2 = \frac{GM}{r_{\rm b}}\left[f^2 U + g^2   \right]\; ,
\end{equation}
where
\begin{equation}
U \equiv 1 + \xi'^2 \, ,
\label{eq:U}
\end{equation}
with the prime indicating a derivative with respect to
$\chi$. Similarly, the gravitational potential can be expressed as 
\begin{equation}
\label{eq:gravpot}
\Phi = -\frac{GM}{r_{\rm b}}\, S\; ,
\end{equation}
where the quantity $S$ is defined by
\begin{equation}
S \equiv (\xi^2 + \chi^2)^{-1/2} \;.
\label{eq:S}
\end{equation}
Furthermore, since we restrict our analysis to `cold' solutions, the
enthalpy term in equation~(\ref{eq:energy}) can be neglected in
comparison with the other terms. Substituting
equations~(\ref{eq:energy}), (\ref{eq:wind9}), (\ref{eq:vsquared})
and~(\ref{eq:gravpot}) into equations~(\ref{eq:energy2})
and~(\ref{eq:l2}) yields
\begin{equation}
\label{eq:energy1}
\varepsilon - \lambda  = \frac{1}{2}(f^2 U + g^2 - 2\xi g) - S =
-\frac{3}{2}\;,
\end{equation}
where we used $\Omega_{\rm B} = (GM/r_{\rm b}^3)^{1/2}$. The numerical value on
the right-hand side of equation~(\ref{eq:energy1}) is obtained by
evaluating this expression at the disk surface; it remains constant
along the flow. 

To make further progress, one can use equation~(\ref{eq:vB}) to write
\begin{equation}
\label{eq:vphi}
g= \frac{v_\phi}{(GM/r_{\rm b})^{1/2}} = \frac{k}{4\pi \rho}
\frac{B_\phi}{(GM/r_{\rm b})^{1/2}} + \xi \, ,
\end{equation}
which, together with equation~(\ref{eq:wind9}), gives
\begin{equation}
\label{eq:bphi}
B_\phi =  \sqrt{\frac{GM}{r_{\rm b}}} k \left[ g - \frac{\lambda}{\xi}
\right] \;.
\end{equation}
Then, substituting equation~(\ref{eq:bphi}) into
quation~(\ref{eq:vphi}), one obtains
\begin{equation}
\label{eq:g}
g = \frac{\xi^2 - m\lambda}{\xi(1 - m)} \; ,
\end{equation}
where $m$ is the square of the Alfv\'en Mach number (the ratio of the poloidal
speed to the poloidal Alfv\'en speed). Finally, substituting
equation~(\ref{eq:g}) into equation~(\ref{eq:energy1}) and using
$\varepsilon - \lambda = -3/2$, one finds
\begin{equation}
T_{\rm w} - f^2 U = \left[\frac{(\lambda - \xi^2) m}{\xi (1 - m)}   \right]^2 \, ,
\label{eq:Tf}
\end{equation}
where
\begin{equation}
T_{\rm w} \equiv \xi^2 + 2S  - 3 \,.
\label{eq:T}
\end{equation}
Note that the gravitational plus centrifugal potential at the position
\{$\xi\, , \,\chi$\} can be expressed as  $-(GM/r_{\rm b})( \xi^2/2 + S)$,
which becomes $-(3/2) GM/r_{\rm b}$ at the base of the wind. It is thus
seen that $-T_{\rm w}/2$ is the gravitational plus centrifugal potential
difference (in units of $GM/r_{\rm b}$) between the point \{$\xi\, ,
\,\chi$\} along a flowline and the base of the flow at \{$\xi=1\, , \,
\chi=0$\}, so that $T_{\rm w}$ must be $\ge 0$ for physical solutions.

One can obtain an expression for $dm/dx \equiv \chi_{\rm A} m'$ 
(where $x \ \equiv \chi/\chi_{\rm A}$ is the vertical coordinate along a
flowline in units of the height of the Alfv\'en point, subscript `A') by
substituting Amp\`ere's Law (equation~\ref{eq:ampere}) into the momentum
equation~(\ref{eq:wind2}) and then combining the vertical component of
equation~(\ref{eq:wind2}) (with the thermal pressure term neglected in
view of the `cold flow' approximation) with the differential form of
equation~(\ref{eq:Tf}). The result, presented by BP82, is
\begin{equation}
\frac{d m}{dx} = \chi_{\rm A} \frac{m S^2 \left[B_{\rm 1}  - B_{\rm
2}(m-1)  - B_{\rm 3} (m-1)^2 \right]}{\xi T_{\rm w} (m-1)(t_{\rm w}-1)}
\;, 
\label{eq:m}
\end{equation}
where 
\begin{equation}
m \equiv \frac{4 \pi \rho (v_r^2 + v_z^2)}{B_r^2 + B_z^2} = \kappa \xi f
J_{\rm w}\, ,
\label{eq:m1}
\end{equation}
\begin{equation}
B_{\rm 1} \equiv 2 m^2 \chi (\xi^2 - \lambda) J_{\rm w}\, ,
\end{equation}
\begin{equation}
B_{\rm 2} \equiv \xi (\chi + \xi \xi') [(5/4)T_{\rm w} + \xi^2 - S]\, ,
\end{equation}
\begin{equation}
B_{\rm 3} \equiv J_{\rm w} \left [\chi (\xi^2 + T_{\rm w}) - f^2 (\chi + \xi
\xi')\right ]\, ,
\end{equation}
\begin{equation}
J_{\rm w} \equiv \xi - \chi \xi'
\label{eq:J}
\end{equation}
and
\begin{equation}
\label{eq:tw}
t_{\rm w} \equiv \frac{1}{T_{\rm w}} \kappa \xi f^3 J_{\rm w}^3 S^2 =
\frac{m^3}{T_{\rm w}} \left(\frac{S}{\kappa \xi}  \right)^2 \, .
\end{equation}
We will also make use of
\begin{equation}
\frac{d \xi}{dx} = \chi_{\rm A} \xi' \, .
\label{eq:xiprime}
\end{equation}

Equation~(\ref{eq:m}) has two singular points. One of them ($m
= 1$) corresponds to the Alfv\'en critical surface, which occurs at the
location ($\chi=\chi_{\rm A}$ or $x = 1$) where the poloidal
velocity component ($v_{\rm p}$) becomes equal to the poloidal component
of the Alfv\'en velocity ($v_{\rm Ap}$). The other singular point
($t_{\rm w}=1$) occurs at the location where $v_{\rm p}$ becomes equal
to the fast-magnetosonic speed and corresponds to the so-called {\em
modified} fast-magnetosonic critical surface
\citep[e.g.][]{B94}. Although an outflow solution that is fully causally
disconnected from the source must cross both of these critical surfaces
\citep[e.g.][]{VTST00}, the condition $t_{\rm w}$ is not expected to
provide any additional constraint on our cold outflow solutions
\citep[see][]{FC04} and we therefore ignore it in this work.

Equations~(\ref{eq:m}) and~(\ref{eq:xiprime}) can be integrated from the
Alfv\'en point to the base of the wind to obtain the wind solution. As
the location of the Alfv\'en critical point ($\chi_{\rm A}$) and the
inclination of the field lines at the base of the wind ($\xib'$) are not
known a priori, we introduce the additional equations
\begin{equation}
\frac{d \chi_{\rm A}}{dx} = 0
\label{eq:xa}
\end{equation}
and
\begin{equation}
\frac{d\xi_{\rm b}'}{dx} = 0 \, ,
\label{eq:xi0}
\end{equation}
so that these quantities can be found self-consistently when the
equations are integrated.

\subsection{Parameters}
\label{subsec:wind_param}

The global, self-similar wind solutions are specified by the following
parameters:
\begin{enumerate}
\item The normalized mass-to-magnetic flux ratio,
\begin{equation}
\kappa \equiv \frac{4 \pi \rho v_{\rm p}}{B_{\rm p}} = \frac{\rhot_{\rm
s}}{a_0^2}\frac{
v_{\rm Kb}}{c_{\rm s}} \;. 
\label{eq:kappa}
\end{equation}

\item The normalized total specific angular momentum,
\begin{equation}
\lambda \equiv \left(r v_\phi - \frac{r B_\phi}{\kappa}\right)
\frac{1}{
r_{\rm b} v_{\rm Kb}} = 1 - \frac{a_0^2}{\rhot_{\rm s}} \frac{c_{\rm
s}}{
v_{\rm Kb}}
b_{\phi \rm b} \,.  
\label{eq:ele}
\end{equation}

\item The inclination of the field lines at the disc surface, measured
by the angle $\varphi$ that the lines make with the vertical,
\begin{equation}
\xi_{\rm b}{'} \equiv \tan \varphi_{\rm b} = b_{r{\rm b}} \,.
\label{eq:incl} 
\end{equation} 
\end{enumerate} 
The rightmost expressions in equations~(\ref{eq:kappa})--(\ref{eq:incl})
show how the global wind parameters can be represented in terms of the
local disc parameters at the base of the wind (see
Section~\ref{subsec:param}). Note that $\kappa (\lambda - 1) =
|b_{\phi{\rm b}}|$ is restricted to a limited range of values for
physically viable solutions (see Fig.~I.2 and WK93). For given choices of two of the above parameters, the third is constrained by the requirement that the solution accelerates past the Alfv\'en critical surface (see Section \ref{subsec:globalsol}). 

\subsection{Boundary Conditions}
\label{subsec:wind_bc}

The system of equations that describes the wind consists of the set of
ODEs given by equations~(\ref{eq:m})
and~(\ref{eq:xiprime})--(\ref{eq:xi0}), together with the algebraic
equations~(\ref{eq:U}), (\ref{eq:S}), (\ref{eq:Tf}), (\ref{eq:T})
and~(\ref{eq:m1})--(\ref{eq:tw}).  This is a two-point boundary value
problem; four boundary conditions are required, which can be specified
either at the Alfv\'en point or at the base of the outflow. They are
applied as follows.
\begin{description}
\item \textbf{At the Alfv\'en point}. Two boundary conditions are
imposed at this location. First, by definition, 
\begin{equation}
m_{\rm A} = 1 \, .
\label{eq:mA}
\end{equation}
This is a 
singular
point of equations~(\ref{eq:Tf}) and~(\ref{eq:m}). Applying the
regularity condition to equation~(\ref{eq:Tf}) yields
\begin{equation}
\xi_{\rm A} = \lambda^{\frac{1}{2}} \, .
\label{eq:LA}
\end{equation}
Note that one can infer from equations~(\ref{eq:mA}) and~(\ref{eq:LA})
that the numerator of equation~(\ref{eq:m}) vanishes identically at the
Alfv\'en point, so no additional boundary condition is obtained by
applying the regularity condition to the latter equation.

Applying l'H\^{o}pital's rule to equations ~(\ref{eq:Tf})
and~(\ref{eq:m}), and using equations~(\ref{eq:mA}) and~(\ref{eq:LA}),
one arrives at the following expressions for $m'_{\rm A}$ and $\xi'_{\rm
A}$:

\begin{equation}
m_{\rm A}' = \frac{2 \xi'_{\rm A}}{(T_{\rm wA} - f_{\rm A}^2 U_{\rm A})^{1/2}}
\label{eq:ma2}
\end{equation}
and
\begin{equation}
C_{\rm 1} m'^{2}_{\rm A} + C_{\rm 2} m'_{\rm A} + C_{\rm 3} = 0 \,,
\label{eq:ma'}
\end{equation}
where
\begin{equation}
C_{\rm 1} \equiv \xi_{\rm A} T_{\rm wA} (t_{\rm wA} - 1) \,,
\end{equation}
\begin{equation}
C_{\rm 2} \equiv \xi_{\rm A} S_{\rm A}^2 (\chi_{\rm A} + \xi_{\rm A}
\xi'_{\rm A}) [(5/4)T_{\rm wA} + \xi_{\rm A}^2 - S_{\rm A}] 
\end{equation}
and
\begin{equation}
C_{\rm 3} \equiv - 4 S_{\rm A}^2 \xi_{\rm A} \xi'_{\rm A} \chi_{\rm A}
J_{\rm wA} \,.
\end{equation}
From equations~(\ref{eq:ma2}) and~(\ref{eq:ma'}) one obtains
$(dm/dx)_{\rm A} = \chi_{\rm A} m_{\rm A}'$ and $(d\xi/dx)_{\rm A} =
\chi_{\rm A} \xi_{\rm A}'$, which can be used to start the integration from the
Alfv\'en point toward the disc.

\item \textbf{At the base of the wind}. Here we could, in principle, use
the condition $\xi_{\rm b} = 1$ from the adopted self-similarity
scaling (see equation~\ref{eq:scaling1}). However, the integration
becomes unstable close to the disc surface and $\xi'$ typically diverges
as $\chi \rightarrow 0$. As a result, a Taylor expansion about \{$\xi =
1\, , \, \chi = 0\}$ is used to obtain the fluid variables at a small
distance ($\delta x \approx 10^{-4}$) above the disc
surface. Specifically, we use
\begin{equation}
f_{\delta x} \approx f_{\rm b}' \delta x + \frac{f_{\rm b}''}{2} \delta x^2
\label{eq:fx}
\end{equation}
and
\begin{equation}
\xi_{\delta x} \approx 1 + \xi_{\rm b}' \delta x + \frac{\xi_{\rm b}''}{2}   \delta x^2
\label{eq:xix}\, ,
\end{equation}
where
\begin{equation}
f_{\rm b}' = \frac{(3 U_{\rm b} - 4)^{\frac{1}{2}} }{[\kappa^2 (\lambda
- 1)^2 + U_{\rm b}]^{\frac{1}{2}}} \; , 
\end{equation}
\begin{equation}
\fb'' = \frac{-9\kappa \xib'^2 - \frac{1}{4}\left (3 D_1 + 5 U_{\rm
b}\right ) \xib' \fb' + D_2}{3 U_{\rm b} - 4} \; , 
\end{equation}
\begin{equation}
\xib'' = -1 -\frac{U_{\rm b}}{4} + \kappa (1 - 3/ \fb'^{2})  \xib' \fb'
- \frac{D_1}{4} \, ,
\end{equation}
\begin{equation}
D_1 \equiv \kappa^2 (\lambda - 9) (\lambda - 1)\, ,
\end{equation}
\begin{equation}
D_2 \equiv \kappa (3 U_{\rm b} - 1) \fb'^2 + 2 \xib' \fb'^3 + 2 \kappa \fb'^4
\end{equation}
and $U_{\rm b} = 1 + \xib'^2$ (see  BP82). The two boundary conditions
applied at $x = \delta x$ are then
\begin{equation}
f(\delta x) = f_{\delta x}
\label{eq:mdx}
\end{equation}
and
\begin{equation}
\xi(\delta x) = \xi_{\delta x}\, ,
\label{eq:xidx}
\end{equation}
where the expressions on the left-hand sides of equations~(\ref{eq:mdx})
and~(\ref{eq:xidx}) are obtained through the integration of the system of
ODEs from the Alfv\'en point toward the disc and those on the right-hand
sides are evaluated using equations~(\ref{eq:fx}) and~(\ref{eq:xix}).
\end{description}

\subsection{Integration of the wind equations}
\label{subsec:wind_int}

 \begin{figure*}
 \centering
 \includegraphics[width=0.8\textwidth]{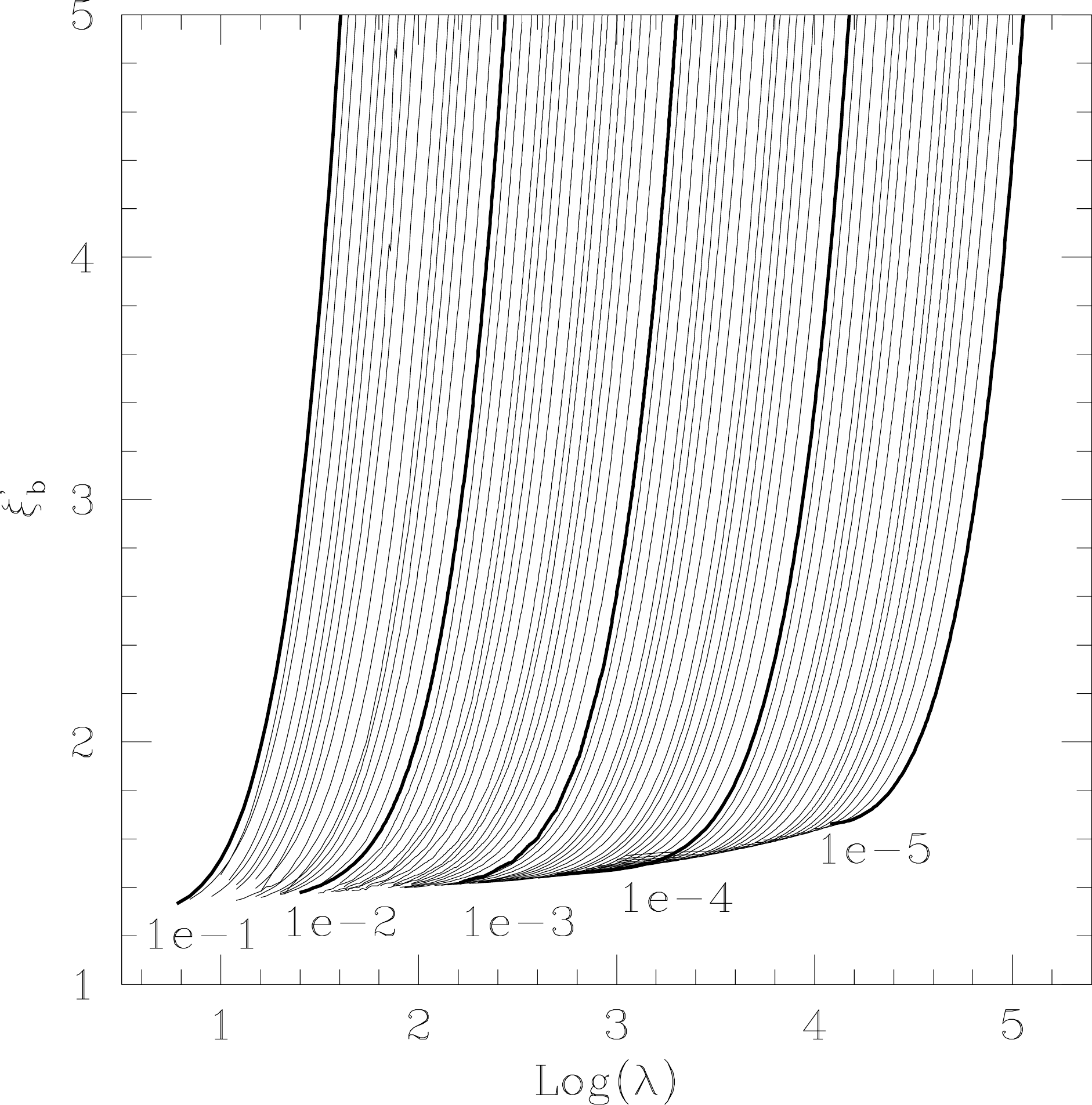} 
\caption{Global, self-similar (BP82-type) wind solutions plotted in the
$\xi_{\rm b}' (\equiv b_{r{\rm b}})$ -- $\log{\lambda}$ 
plane, where
$\lambda$ is the normalized total specific angular momentum (including
the matter and magnetic field contributions) and $\xi'_{\rm b}$
measures the inclination of the field lines at the base of the wind. The curves
are labelled by the normalized mass-to-magnetic flux ratio, $\kappa$: the
values of $\kappa$ that correspond to the darker curves are indicated in
the figure. These solutions are available electronically in tabular
form, as detailed in the text.}
\label{fig:windres}
\end{figure*}

To integrate the wind equations, it is first necessary to choose the
values of the free parameters $\kappa$ and $\lambda$ and supply initial
(guessed) values of $\chi_{\rm A}$ and $\xib'$. In the coupled
disc--wind solutions considered here, the adopted values of $\kappa$ and
$\lambda$ -- as well as the initial estimate of $\xib'$ -- are, in fact,
calculated from the rightmost expressions in
equations~(\ref{eq:kappa})--(\ref{eq:incl}), using the values of the
fluid variables at the base of the wind that are obtained at the end of
the iterations on the disc solution (see Section~\ref{subsec:globalsol}).
One can then use equations~(\ref{eq:ma2}) and~(\ref{eq:ma'}) to evaluate
$(d \xi/ d x)_{\rm A} = \chi_{\rm A} \xi'_{\rm A}$ and $(d m/ d x)_{\rm
A} = \chi_{\rm A} m'_{\rm A}$ and start the integration of
equations~(\ref{eq:m}) and~(\ref{eq:xiprime}) from the Alfv\'en point
($x_{\rm A} = 1$) toward the base of the wind.

Below the Alfv\'en point (i.e. for $x < 1$), the value of $\xi'$ on the
right-hand side of equation~(\ref{eq:xiprime}) is obtained by
substituting $U$, $S$, $T_{\rm w}$, $f$ and $J_{\rm w}$ (which are
found from equations~\ref{eq:U}, \ref{eq:S}, \ref{eq:T}, \ref{eq:m1}
and \ref{eq:J}, respectively), as well as $m$ and $\xi$ (which are found
from the integrals of equations~\ref{eq:m} and~\ref{eq:xiprime}) into
equation~(\ref{eq:Tf}). This yields the following quadratic equation
for $\xi'$:
\begin{equation}
(1 - K_1 \chi_{\rm A}^2 x^2) \xi'^2 + 2 \chi_{\rm A} x K_1 \xi + (1 - K_1
\xi^2) =0 \, ,
\end{equation} 
where 
\begin{equation} 
K_1 \equiv \left(\frac{\kappa \xi}{m} \right)^2 \left \{T_w -
\left[\frac{(\lambda - \xi^2)m}{\xi(1 - m)} \right]^2 \right \} \; .  
\end{equation} 
Out of the two possible roots of this equation, we choose the one that
satisfies the condition ${\rm tan} \ \varphi \equiv \xi' < {\rm tan}\
\theta \equiv \xi/\chi$, as appropriate for a solution that describes a
collimating wind.

As discussed at the end of Section~\ref{subsec:wind_bc}, the boundary
conditions at the disc are actually applied at a small distance $\delta
x$ above the surface, where the values derived by integrating down from
the Alfv\'en surface are matched (via equations~\ref{eq:mdx}
and~\ref{eq:xidx}) to the values obtained by stepping off the mid-plane
with a Taylor expansion. The full solution is then found by iterating on
$\chi_{\rm A}$ and $\xib'$ until convergence is reached. Using this
procedure, we derived wind solutions for a wide range of values of the
parameters $\kappa$ and $\lambda$ (see Fig.~\ref{fig:windres}).  This
solution `library' is useful when one proceeds to smoothly match a
radially localized disc solution to a global wind solution (see
Section~\ref{subsec:globalsol}). It is available, in tabular form, on
the VizieR data base of astronomical catalogues
(http://cdsarc.u-strasbg.fr/).

\subsection{Matching the (localized) disc and (global) wind solutions}
\label{subsec:globalsol}

Having derived a localized disc solution as described in
Section~\ref{subsec:discsol}, one calculates the associated parameters
of a self-similar wind solution ($\kappa$, $\lambda$ and $\xi'_{\rm b}$)
using the rightmost expressions in
equations~(\ref{eq:kappa})--(\ref{eq:incl}). The parameter combination
obtained in this way does not, however, generally correspond to a wind
solution that crosses the Alfv\'en critical surface. In the next step,
one derives a self-similar wind solution as described in
Section~\ref{subsec:wind_int}, using the given values of $\kappa$ and
$\lambda$ and employing the value of $\xib'$ from
equation~(\ref{eq:incl}) as an initial guess in implementing the
boundary conditions at the base of the wind. The final value of
$\xi_{\rm b}'$ from the wind iteration will, in general, be different
from $b_{r{\rm b}}$, so that equation~(\ref{eq:incl}) will not be
satisfied. To obtain a self-consistent disc--wind solution, one iterates
on the disc and the wind solutions, using $\epsilon$ (the normalized
mid-plane radial velocity) as an adjustable parameter, until the value
of $\xi'_{\rm b}$ from the wind iteration satisfies
equation~(\ref{eq:incl}). The CPU time to compute either a disc or a
wind solution as described above is typically under a few seconds on a
2.4 GHz, AMD Opteron system. On the other hand, the CPU time associated
with obtaining a matched disc--wind solution varies with the particular
case, but it is not unusual for the entire procedure to take up to 15 --
20 minutes.

\section{Results}
\label{sec:results}

In this section we discuss the main features of radially localized
solutions of well-coupled, wind-driving discs in the Hall diffusivity
domain.  In Paper~I we identified four parameter sub-regimes in this
domain by imposing general physical constraints on viable solutions in
the context of the {\it hydrostatic approximation}, wherein the vertical
velocity component is neglected. This approximation has made it possible
to derive algebraic relations that characterize the extent of each
sub-regime and the distinguishing properties of the associated
solutions. These are summarized in Tables~\ref{table:constraints}
and~\ref{table:constraints1}, respectively, of Appendix~\ref{sec:appA},
where we also collect some of the algebraic expressions that were
employed in the derivation of these results. In
Section~\ref{subsec:illus} we present representative solutions for these
four sub-regimes (labelled by the Roman numerals i through iv) and
compare their properties with the predictions of
Table~\ref{table:constraints1}. We then proceed
(Section~\ref{subsec:test}) to analyse the dependence of the solutions
on the conductivity component ratio $|\sigma_{\rm H}|/\sigma_\perp$ as
well as on the sign of the Hall conductivity $\sigH$ (i.e. on the
magnetic field polarity) and on the parameter $\epsilon$. We also test
whether physically viable solutions are indeed excluded from the
parameter regimes that are `forbidden' according to the hydrostatic
analysis.  Finally, in Section~\ref{subsec:global}, we present
illustrative disc solutions that are matched to self-similar wind
solutions and briefly discuss the properties of the joint disc--wind
solutions obtained in this way.

In all of our disc models we set the parameter $\epsilon_{\rm B}$ to be
identically zero. This parameter, which measures the radial drift
velocity of the poloidal magnetic field lines (see Section~I.3.13),
depends on the global distribution of $B_z$ along the disc and,
therefore, cannot be obtained self-consistently in our localized
formulation. The justification for setting $\epsilon_{\rm B} = 0$ in our
analysis is discussed in WK93 and in Appendix~I.A. Basically, solutions
characterized by the same value of the parameter combination ($\epsilon
- \epsilon_{\rm B}$) are qualitatively similar, a property that results
from the fact that the only change in the underlying system of equations
brought about by switching to a reference frame anchored in the radially
drifting poloidal flux surfaces (which move with the radial velocity
$v_{\rm Br0} = - \epsilon_{\rm B} c_{\rm s}$) is the substitution of
$v_r-v_{\rm Br0}$ for the radial velocity $v_r$ in all the equations
except the angular momentum conservation relation, which remains
unchanged.
 
An additional simplification that we employ is to have the Ohm, Pedersen
and Hall conductivity terms ($\sigma_{\rm O}$, $\sigma_{\rm P}$ and
$\sigma_{\rm H}$, respectively) scale with the gas density
$\rho$ and magnetic field amplitude $B$ as $\rho/B^2$. 
We adopt this dependence since it results in the local
matter--field coupling parameter $\Lambda$ being constant with height.
This is convenient for comparison of our solutions with the
analytic results of Paper~I, which were
effectively obtained 
under the same approximation, as well as with the solutions derived by
WK93 in the ambipolar diffusivity regime, where the corresponding value
of $\Lambda$ was similarly assumed not to vary with height. Further
calculations, exploring the properties of solutions computed with more
realistic ionization and conductivity profiles, will be presented
elsewhere.

\subsection{Illustrative disc solutions in the Hall parameter sub-regimes}
\label{subsec:illus}

\begin{figure*}
\begin{minipage}{1\textwidth}
\centering
\includegraphics[height=10cm]{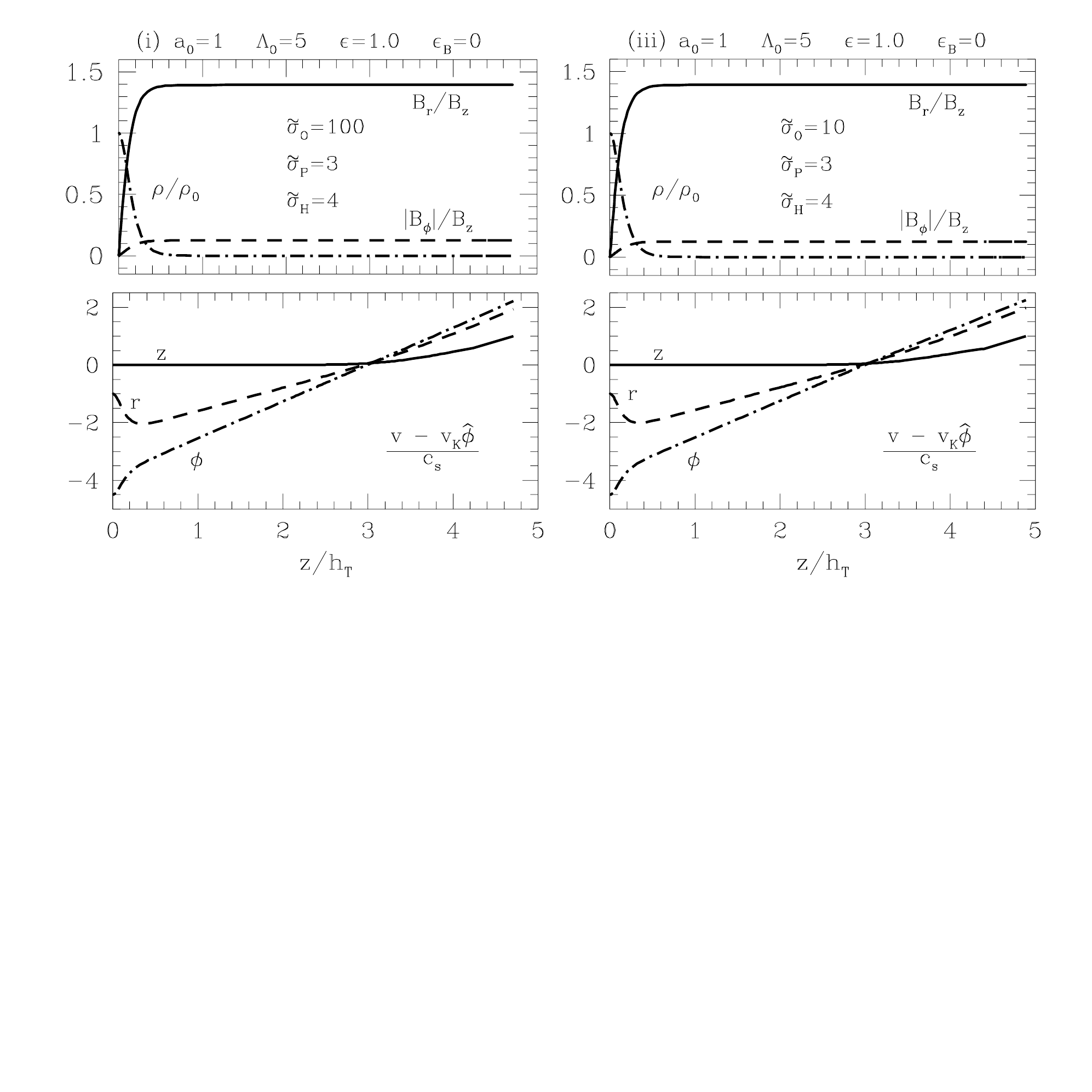}
\caption{Structure of illustrative disc solutions as a
function of height above the midplane (measured in units of the tidal
scaleheight $h_{\rm T}$) for the Hall conductivity sub-regimes labelled in
Table~\ref{table:constraints} as Case~(i) (left panels) and Case~(iii)
(right panels) and the model parameters shown in the figure 
(where all the listed conductivity values pertain to the mid-plane). 
For
these two cases, the matter--field coupling parameter $\lao > 1/2$. 
In both solutions, the top panel shows the gas density
(normalized by the midplane value) as well as the radial and azimuthal
components of the magnetic field (normalized by the vertical component,
which is constant with height). The bottom panel depicts the velocity
components in a frame rotating at the Keplerian speed, normalized by
the isothermal sound speed $c_{\rm s}$.  The curves terminate at the
sonic point. The key properties of both solutions are listed in
Table~\ref{table:boundary}.}
\label{fig:i-ii}
\vspace{0.3cm}
\end{minipage}
\hspace{0.06\textwidth}
\begin{minipage}{1\textwidth}
\centering
\includegraphics[height=10cm]{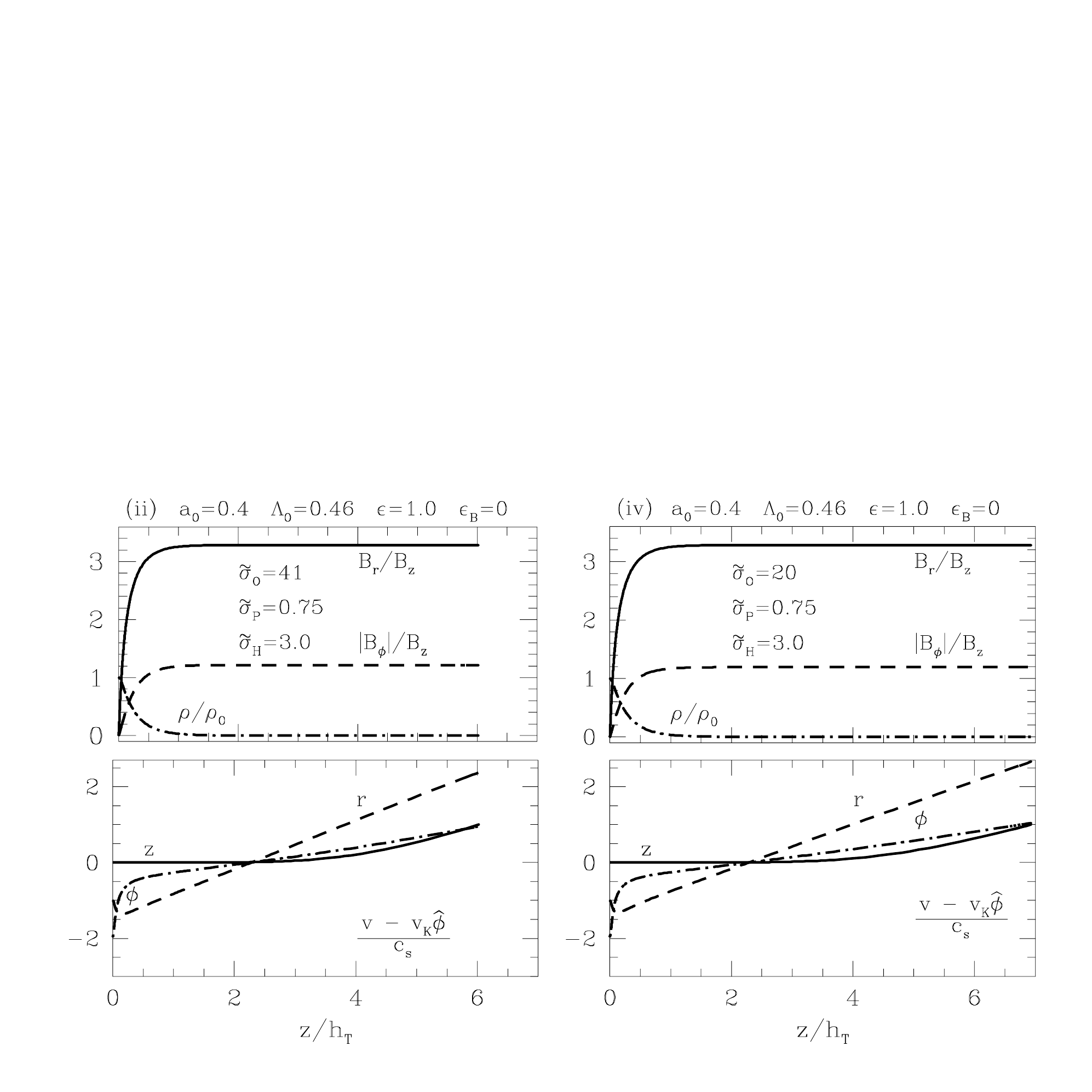}
\caption{Same as Fig.~\ref{fig:i-ii} except that the depicted solutions
correspond to Cases~(ii) (left panels) and~(iv) (right panels) of
Table~\ref{table:constraints}, for which $\lao < 1/2$. Key properties of
these solutions are summarized in Table~\ref{table:boundary}.}
\label{fig:iii-iv}
\end{minipage}
\end{figure*}

\begin{table*}
\caption{Key properties of the solutions shown in Figs.~\ref{fig:i-ii}
and \ref{fig:iii-iv}. We list the 1st--3rd inequalities of Table
\ref{table:constraints} (representing, respectively, the requirement for
a sub-Keplerian flow within the disc, the wind-launching condition and
the lower bound on the height of the base of the wind) for each case as
well as the mid-plane values of 
the ion slip factor $s\equiv \bete \beti$ and of the matter--field
coupling parameter $\Lambda = a^2 \, \tilde{\sigma}_\perp$, which is
equal to $\Upsilon|\beti|$ in the limit $(|\sigH|/\tilde{\sigma}_\perp -
1) \ll 1$ (see equation~I.95). These parameters form the basis of the
classification scheme of viable solutions in the Hall diffusivity regime
(see Section I.5). The base of the wind and the sonic-surface heights
($z_{\rm b}$ and $z_{\rm s}$, respectively) are in units of both the
tidal scaleheight $h_{\rm T}$ and the actual (magnetically reduced)
pressure scaleheight $h$, where the latter is determined in each case
as the value of $z$ where the density drops to $\rho_0/\sqrt{e}$.  The
mass accretion rate per disc circumference is evaluated from $\dot
M_{\rm in}/2\pi r_0 = - 2 \int_0^{z_{\rm b}}{\rho v_r dz}$, and the
listed value ($\dot{\cal{M}}_{\rm in}$) is this quantity normalized by
$\rho_0 c_{\rm s} h_{\rm T}$.}

\begin{tabular}{ccccc}
\hline
Illustrative & Case (i) & Case (iii) & Case (ii) & Case (iv) \\
disc solution & Fig.~2 (left) & Fig.~2 (right) & Fig.~3 (left) & Fig.~3
(right) \\
\hline
& & & & \\

Constraints & $\frac{1}{\sqrt{2\Upsilon_0}} \lesssim a_0 \lesssim 2
\lesssim \epsilon\Upsilon_0$ & $\frac{1}{\sqrt{2\Upsilon_0}} \lesssim
a_0 \lesssim 2 \lesssim \epsilon \Upsilon_0 s_{\rm 0}$ &
$\sqrt{\beta_{\rm i0}} \lesssim a_0 \lesssim 2
\sqrt{\Upsilon_0 \beta_{\rm i0}} \lesssim \frac{\epsilon}{2\beta_{\rm
i0}}$ & $\sqrt{\beta_{\rm i0}} \lesssim a_0 \lesssim 2\sqrt{\Upsilon_0
\beta_{\rm i0}} \lesssim \frac{\epsilon \beta_{\rm e0}}{2}$ \\

& & & & \\

& $0.2 \lesssim 1 \lesssim 2 \lesssim 9$ & 
$0.1 \lesssim 1 \lesssim 2 \lesssim 10$ & 
$0.4 \lesssim 0.4 \lesssim 1 \lesssim 3$ & 
$0.3 \lesssim 0.4 \lesssim 1 \lesssim 3$  \\ 

& & & & \\

 $s_{\rm 0} \equiv \beteo \beta_{\rm i0}$ & $11$ & $0.2$ &  $2.2$ & $0.6$ \\
 
$\lao \equiv a_{\rm 0}^2 \ \tilde{\sigma}_{\perp{\rm 0}}$
 & $5.0$ & $5.0$ & $0.46$ & $0.46$ \\

& & & & \\

Properties & & & & \\

& & & & \\

$\sigpar$ & $100$ & $10.0$ &  $41$ & $20$ \\
$\sigH$ & $4.0$ & $4.0$ &  $3.0$ & $3.0$ \\
$\sigP$ & $3.0$ & $3.0$ &  $0.75$ & $0.75$ \\
$\beta_{\rm e0}$ & $16$ & $1.6$ &  $12.9$ & $6.3$ \\
$\betio$ & $0.69$ & $0.13$ &  $0.17$ & $0.1$ \\
$\Lambda_{\rm 0}$ & $5.0$ & $5.0$ &  $0.46$ & $0.46$ \\
$\upo$ & $9.1$ &
$50$ &  $2.8$ & $5.2$ \\
$a_o$ & $1.0$ & $1.0$ & $0.4$ & $0.4$ \\
$\epsilon$ & $1.0$ & $1.0$ & $1.0$ & $1.0$ \\
$h/h_{\rm T}$ & $0.12$ & $0.12$ & $0.16$ & $0.16$\\
$z_{\rm b}/h_{\rm T}$ & $3.0$ & $3.0$ & $2.3$ & $2.3$\\
$z_{\rm s}/h_{\rm T}$ & $4.7$ & 
$4.9$ & $6.0$ & $6.9$\\    
$z_{\rm b}/h$ & $25$ &  
$25$ & $14$ & $14$\\
$z_{\rm s}/h$ & $39$ & 
$41$ & $37$ & $42$\\    
$\rho_s/\rho_0$ & $2.1\ee{-10} $ & 
$2.2\ee{-10} $ & $1.1 \ee{-5}$ & $5.9\ee{-6}$\\
$\xi_{\rm b}^{\prime} \equiv B_{r,{\rm b}}/B_z$ & $1.4$ & $1.4$ & $3.3$
& $3.3$\\
$|B_{\phi,b}|/B_z$ & $0.12$ & 
$0.13$ & $1.2$ & $1.2$\\
$\dot{\cal{M}}_{\rm in}$ & $0.48$ & $0.48$ & $0.72$ & $0.70$\\   
\hline                                
\end{tabular}
	\label{table:boundary}
\end{table*}

In this subsection we present representative solutions for the four Hall
parameter sub-regimes 
(see Table~\ref{table:boundary}).
 We first divide them according to whether the value of the mid-plane
Elsasser number $\lao$ is $>1/2$ (Cases~i and~iii; Fig.~\ref{fig:i-ii})
or $<1/2$ (Cases~ii and~iv; Fig.~\ref{fig:iii-iv}), which is one of the
two classification criteria that define these
sub-regimes.\footnote{Computing solutions for $\lao < 1/2$ has proven to be
numerically challenging. For this reason, the solutions presented in
Fig.~\ref{fig:iii-iv} correspond to values of $\lao$ that are only
slightly smaller than 1/2.}
We further
divide them according to the second classification criterion, which is
whether $s_{\rm 0} \equiv \beta_{\rm e0} \beta_{\rm i0}$, the mid-plane
value of the ion slip factor, is $> 1$ (Cases~i and~ii, shown in the
left panels of Figs.~\ref{fig:i-ii} and~\ref{fig:iii-iv}, respectively)
or $<1$ (Cases~iii and~iv, depicted in the corresponding right panels of
these figures). The variables $\betio$ and $\beteo$ are, respectively,
the mid-plane values of the ion and electron Hall parameters; see
equation~\ref{eq:bebiq}. Now, the hydrostatic analysis implies that
viable solutions must satisfy $s_0 > 1$ in the ambipolar diffusivity
regime (WK93) and $s_0<1$ in the Ohm diffusivity regime (Paper~I). One
can therefore classify Cases~(i) and~(ii) as being in the `ambipolar
diffusion-modified' Hall regime and Cases~(iii) and~(iv) as being in the
`Ohm diffusion-modified' Hall regime. This description is supported by
the fact that the parameter constraints and solution properties derived
for the Hall Case~(i) are identical to those in the ambipolar
diffusivity limit, with a similar correspondence holding between the
Hall sub-regime~(iii) and the Ohm sub-regime~(i) (see Paper~I).

We use the solution for Case~(i) (left panel of Fig.~\ref{fig:i-ii}) as
a representative model that illustrates the overall properties of
well-coupled wind-driving discs. The top panel of this figure shows the
gas density as well as the radial and azimuthal components of the
magnetic field as functions of height between the midplane and the sonic
point.  The bottom panel depicts the velocity components. This layout is
used also in all the other figures in this paper that depict the
vertical structure of the disc solutions.

The representative disc solution exhibits the following overall
properties (WK93). In the \emph{quasi-hydrostatic layer} straddling the
mid-plane ($0 \lesssim |\tilde z| \lesssim 0.4$ in this example), the
magnetic field lines are radially bent and azimuthally sheared (see
equations~\ref{eq:phi_ampered} and \ref{eq:r_ampered}), and matter loses
angular momentum to the magnetic field. This enables the disc material
to flow toward the protostar (i.e. $\epsilon \equiv -w_{r{\rm 0}} > 0$)
and results in a radial drag on the ionized disc component (assuming
that $\epsilon_{\rm B}$ remains $<\epsilon$; see WK93 and
Appendix~I.A). This drag must be balanced by magnetic tension (see
equations~\ref{eq:phi_ampered}, \ref{eq:E-E'1} and~\ref{eq:jphi}), which
is consistent with the field lines bending radially
outward. Furthermore, as the magnetic tension also partially supports
the fluid against the gravitational pull of the protostar (see
equations~\ref{eq:r_motiond} and~\ref{eq:phi_ampered}), the gas motion
in this region is sub-Keplerian.  Further up but still below the base of
the wind (defined as the height $\tilde z_{\rm b}$ where $v_\phi =
v_{\rm K}$ and located in this solution at $ z \approx 3.0$) lies the
\emph{transition zone} ($0.4 \lesssim |\tilde z| \lesssim 3.0$). In this
layer the magnetic pressure comes to dominante the thermal pressure
because of the strong drop in the gas density away from the midplane,
and the magnetic field lines are locally straight. Note that everywhere
within the disc the radial velocity is negative, the azimuthal velocity
is sub-Keplerian, and the vertical velocity is subsonic (highly so in
the quasi-hydrostatic layer). Finally, as the angular velocity of the
magnetic field lines is nearly constant along the poloidal flux surfaces
(it is exactly constant only when $\epsilon_{\rm B}$ is identically
zero; see \citealt{K89}), their azimuthal velocity increases with radius
until, eventually, they overtake the fluid (whose azimuthal velocity
decreases with radius according to the Keplerian rotation law). In the
hydrostatic limit ($\vz \rightarrow 0$) and with $\epsilon_{\rm B} = 0$
this occurs at the height $\tilde z_{\rm b}$, and under the same
assumptions this is also the location where $v_r$ changes
sign.\footnote{In an exact solution, where $\vz$ is $>0$ near the top of
the disc, the heights where $\vf =\vk$, $\vf = r\Omega_{\rm B}$ and $\vr
= 0$ do not exactly coincide, although typically they remain close to each
other.} The solution region above $\tilde z_{\rm b}$ (between $\tilde
z_{\rm b} \approx 3.0$ and the sonic surface at $\tilde z_{\rm s}
\approx 4.7$) represents the \emph{base of the wind}.  In this region
$\vr$ is $>0$, $\vf$ is super-Keplerian and the magnetic field transfers
angular momentum back to the matter, causing the uppermost layers of the
disc to be driven out centrifugally (with the mass flux in the wind
effectively fixed by the regularity condition at the sonic critical
point of the disc solution).

The distinguishing properties of the four Hall sub-regimes, as inferred
in the context of the hydrostatic approximation, are listed in
Table~\ref{table:constraints1}. To see how the expressions derived in
this approximation for $|db_r/db_\phi|_0$, $\tilde h \equiv h/h_{\rm T}$
(where $h$ is the magnetically reduced density scaleheight, defined as the
location where the density drops to $\rho_{\rm 0}/\sqrt{e}$) and $\tilde z_{\rm
b}/\tilde h$ compare with the properties of the full solutions listed in
Table~\ref{table:boundary}, we need to relate the parameters
$\betio$, $\beteo$ and $\upo$ used in the two-component plasma
formulation of Paper~I to the tensor conductivity components employed in
this work. The parameter $\upo$ is the mid-plane value of the
neutral--ion coupling function given by equation~\ref{eq:eta} and
is equal to the Elsasser number in the ambipolar diffusivity limit.
The relationships between these variables and the conductivity ratios
can be obtained from equations~I.95--I.97 and reduce, in the limit
$q\equiv \beta_{\rm i}/\beta_{\rm e} \ll 1$, to
\begin{equation} 
\upo \approx \lao \left(\frac{\sigP}{\tilde
\sigma_\perp} - \frac{\tilde \sigma_\perp}{\sigpar} \right)^{-1}\ ,
\label{eq:upo} 
\end{equation} 
\begin{equation} 
\betio \approx
\left(\frac{\sigP}{\tilde\sigma_\perp} -
\frac{\tilde\sigma_\perp}{\sigpar} \right) \left(
\frac{\sigH}{\tilde\sigma_\perp} \right)^{-1} 
\label{eq:betio}
\end{equation} 
and 
\begin{equation} 
\beteo \approx \frac{\sigH \sigpar}{{\tilde \sigma_\perp}^2}\ .  
\label{eq:beteo} 
\end{equation}

\subsection{Parameter dependence of the solutions}
\label{subsec:test}
\subsubsection{Dependence on the ratio $|\sigH|/\tilde \sigma_\perp$}
\label{subsubsec:sigH}

\begin{figure}%  figure placement: here, top, bottom, or page
   \centering
   \includegraphics[width=0.45\textwidth]{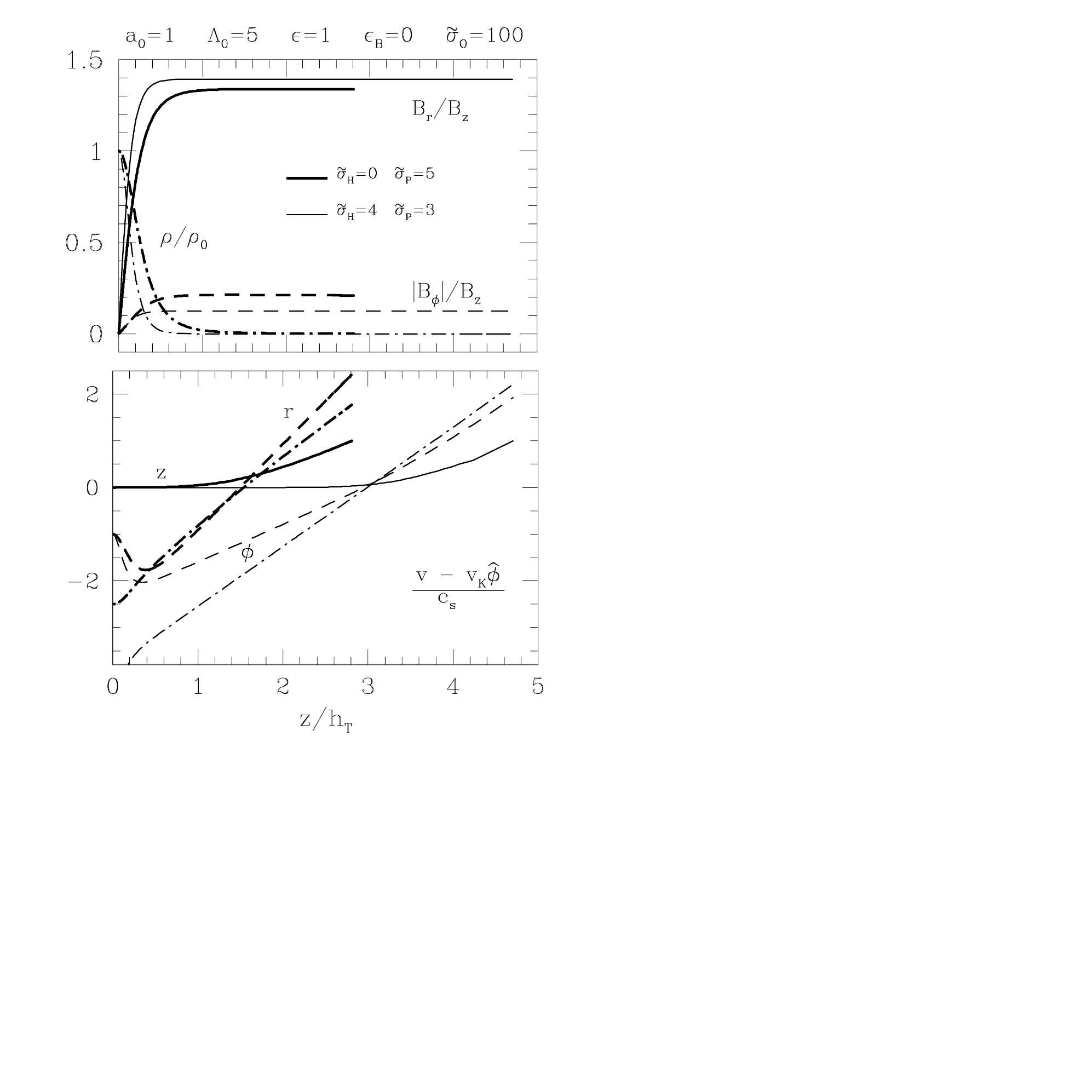} 
   \caption{
Structure of a wind-driving disc solution as a function of $z/h_{\rm T}$
for the disc parameters shown in the figure and two different
configurations of the conductivity tensor 
(whose listed components all pertain to the mid-plane).  
The thick lines correspond to
the ambipolar diffusivity limit (for which $\sigH = 0$ and $\sigpar \gg
\sigP$) and the thin lines to a case where $\sigH > \sigP$.
The normalized upward mass flux
is $\rhot w_z = \rhot_{\rm s} = 1.1 \ee{-3}$ for the solution in the
ambipolar diffusivity limit and $1.4 \ee{-6}$ in the Hall case.}

   \label{fig:10_1}
\end{figure}

\begin{figure}%  figure placement: here, top, bottom, or page
   \centering
   \includegraphics[width=0.5\textwidth]{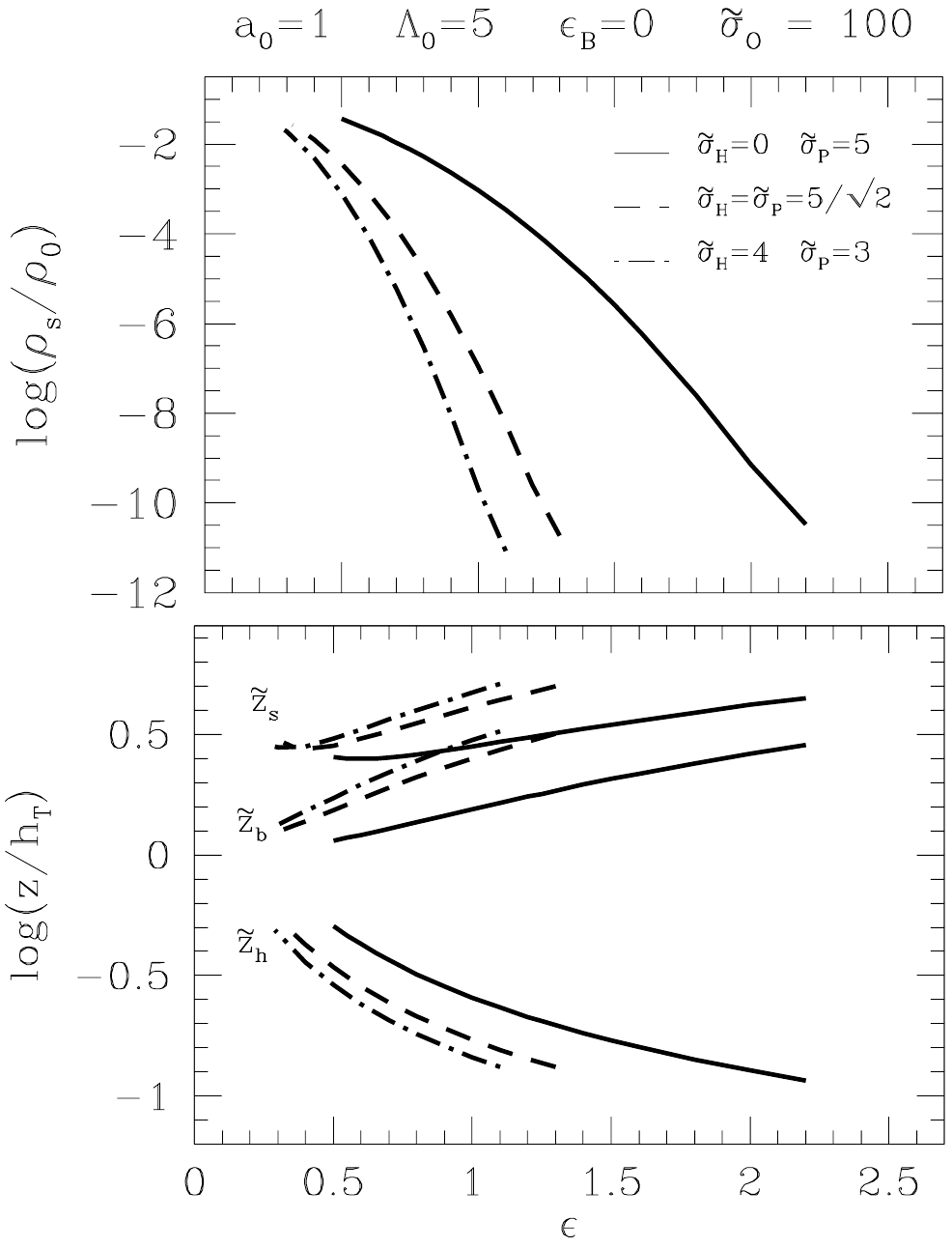} 
\caption{
\emph{Top panel:} Density at the sonic point ($\rhot_{\rm s} \equiv \rho_{\rm
   s}/\rho_{\rm 0}$)
as a function of the normalized inward velocity at the midplane
($\epsilon$) for three configurations of the conductivity tensor
characterized by the same values of $\sigpar$ and $\tilde\sigma_\perp$
but different values of the ratio $\sigH/\tilde\sigma_\perp$. 
(All the listed conductivities pertain to the mid-plane.) 
The
other model parameters have the same values (indicated in the figure)
in each case. \emph{Bottom panel:} As above, but now plotting
the vertical location of the sonic point ($\tilde z_{\rm s}$), the
base of the wind ($\tilde z_{\rm b}$) and the magnetically reduced
scaleheight ($\tilde z_{\rm h} \equiv \tilde{h}$). In both panels, the solid 
lines show the ambipolar diffusivity limit ($\sigH = 0$), the dashed
lines represent the case where $\sigH = \sigP$ and the dot-dashed lines
depict a case where $\sigH > \sigP$.}
   \label{fig:5_10}
   \end{figure}

The conductivity ratio $\sigH/\tilde \sigma_\perp$ serves to distinguish
the ambipolar diffusivity domain ($\sigpar \gg \sigP \gg |\sigH|$) from
the Hall diffusivity regime ($\sigP \ll |\sigH| \ll \sigpar$): the
ambipolar diffusivity limit corresponds to this ratio tending to 0,
whereas the Hall limit corresponds to this ratio tending to 1. To
illustrate the difference between these two regimes, it may be
instructive to compare a solution in the ambipolar diffusivity regime,
which satisfies $\lao \gtrsim 1$ and $s_0 > 1$ (see Paper~I), with a
solution in the Hall sub-regime~(i), which, as we already noted in
Section~\ref{subsec:illus}, satisfies the same inequalities and, in
fact, identical parameter constraints (first line in
Table~\ref{table:constraints}). We show such a comparison in
Figs.~\ref{fig:10_1} and~\ref{fig:5_10}. Fig.~\ref{fig:10_1} depicts the
vertical run of the density as well as of the transverse magnetic field
and the velocity components for two representative solutions in the
above-mentioned diffusivity regimes for the same values of the
parameters $\lao$, $a_0$ and $\epsilon$. Fig.~\ref{fig:5_10}, in turn,
shows the dependence of the density decrement $\rho_{\rm s}/\rho_0$ at
the sonic surface and of the characteristic heights $\zt_{\rm h} \equiv
\tilde h$, $\zt_{\rm b}$ and $\zt_{\rm s}$ on the parameter $\epsilon$
for the same two choices of the conductivity components (as well as for
a solution with an intermediate value of $\sigH/\tilde \sigma_\perp =
5/\sqrt{2}$) and the same values of $\lao$ and $a_0$ as in
Fig.~\ref{fig:10_1}.

The most noticeable differences between the two solutions shown in
Fig.~\ref{fig:10_1} are the faster change of the flow variables
(corresponding to a smaller value of $\tilde h$) and the larger vertical
extent of the displayed solution (corresponding to a higher value of
$\zt_{\rm s}$) in the Hall case. The higher value of $\zt_{\rm s}$
implies a lower density at the sonic surface and therefore a lower
normalized upward mass flux ($=\rhot_{\rm s}$) for the Hall solution
(see Fig.~\ref{fig:5_10}). The decrease (increase) of $\zt_{\rm h}$ 
($\zt_{\rm s}$) with increasing
$|\sigH|/\tilde\sigma_\perp$ when all the other parameter values are
held constant is also evident from the curves plotted in
Fig.~\ref{fig:5_10}. The latter figure further demonstrates that the
behaviour of $\zt_{\rm s}$ is similar to that of $\zt_{\rm b}$, which
suggests that $\zt_{\rm b}$ could be used as a proxy for $\zt_{\rm s}$
in analysing this trend. The parameter dependence of $\tilde h$ and
$\zt_{\rm b}$ can be estimated using the hydrostatic approximation, as
discussed in Paper~I, and is given by equations~(\ref{eq:ht})
and~(\ref{eq:zb}), respectively. Given that $\lao \rightarrow
\upo\betiabs$ as $|\sigH|/\tilde\sigma_\perp \rightarrow 1$ and
assuming that $\lao$, $a_0$ and $\epsilon$ are held constant, 
the above equations imply
that, to leading order in $\betiabs$, $\tilde h \propto \betiabs$ and
$\zt_{\rm b} \propto 1/\betiabs$ in the ambipolar regime
and in the Hall sub-regimes~(i) and~(ii),
for which $s_0 = \beteo\betio 
\gg 1$. On the other hand,
equation~(\ref{eq:betio}) implies, given that $\tilde\sigma_\perp =
\lao/a_0^2$ (equation~I.87) and $\sigpar$ are fixed, that $\betiabs$
\emph{decreases} as $|\sigH|/\tilde\sigma_\perp$ goes up (and
$\sigP /\tilde\sigma_\perp = (\tilde\sigma_\perp^2 -
\sigH^2)^{1/2}/\tilde\sigma_\perp$ goes down). Taken together, these
results explain the manifested dependence of the solutions in
Figs.~\ref{fig:10_1} and~\ref{fig:5_10} on $|\sigH|/\tilde\sigma_\perp$.

As a further check on the applicability of the analytical framework
developed in Paper~I, we note that equations~(\ref{eq:ht})
and~(\ref{eq:zb}) imply 
a similar
dependence of $\zt_{\rm h}$ and $\zt_{\rm b}$ on
$|\sigH|/\tilde\sigma_\perp$ in the limit $s_0 \ll 1$ (i.e. in the Ohm
diffusion-modified Hall regime; Cases~iii and~iv). In fact, fixing the
values of the same parameters as before, these two equations imply that,
to leading order in small ratios, $\tilde h \propto 1/|\beteo|$ and
$\zt_{\rm b} \propto |\beteo|$ in this case, while
equation~(\ref{eq:beteo}) shows that $|\beteo|$ \emph{increases} as
$|\sigH|/\tilde\sigma_\perp$ goes up (assuming again that
$\tilde\sigma_\perp$ and $\sigpar$ remain unchanged). We have verified
that solutions in these regimes indeed exhibit the expected dependence
on $|\sigH|/\tilde\sigma_\perp$.

\begin{figure}%  figure placement: here, top, bottom, or page
   \centering
   \includegraphics[width=0.45\textwidth]{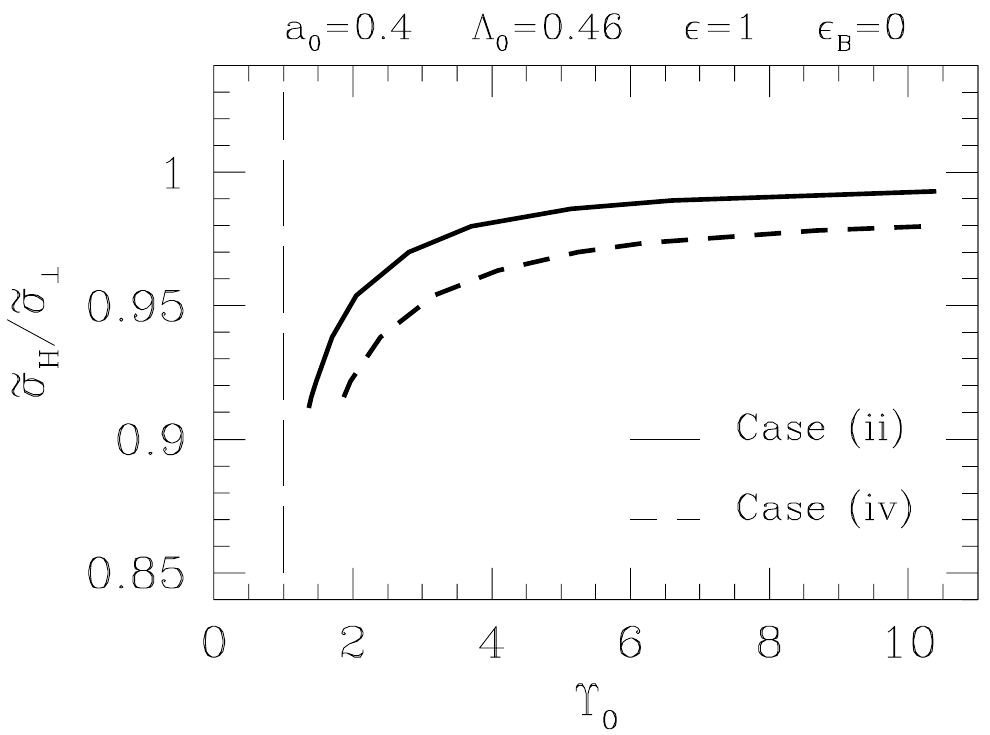} 
\caption{Dependence of the ratio $\sigH/\tilde\sigma_\perp$ on the
parameter $\upo$ for solutions corresponding to the indicated parameter
values. The solid and short-dashed curves show solutions for the Hall
sub-regimes~(ii) (computed with $\sigpar = 41$ at the mid-plane) and~(iv)
(computed with $\sigpar=20$ at the mid-plane), respectively. In both
cases, $\tilde\sigma_{\perp 0} = 3.1$. All viable solutions lie to the
right of the vertical long-dashed line, where $\upo > 1$, as predicted
by the hydrostatic analysis.
}
   \label{fig:eta_sh}
\end{figure}

Yet another test of the predictions of the hydrostatic analysis
regarding the dependence of the solutions on
$|\sigH|/\tilde\sigma_\perp$ can be constructed using
equation~(\ref{eq:upo}) for the parameter $\upo$. As discussed in
Section~I.6 and illustrated in Fig.~I.2, this analysis indicates that
the requirement $\upo \gtrsim 1$ (i.e. that the mid-plane neutral--ion
momentum exchange time be shorter than the local orbital time) is a
fundamental constraint on viable wind-driving disc models of the type
that we consider, and applies to solutions in all the diffusivity regimes
(ambipolar, Hall and Ohm). This condition follows directly from the
inequalities $\upo \betiabs > 1/2$ and $\betiabs < 1$ that characterize
the Hall Cases~(i) and~(iii), but, as can be seen by combining the first
two parameter constraints reproduced in the first row of
Table~\ref{table:boundary}, it can be formally inferred (within the
framework of the hydrostatic analysis) to apply also in the other Hall
sub-regimes. The illustrative solutions for the Cases~(ii) and~(iv) in
Fig.~\ref{fig:iii-iv} satisfy this inequality, as verified explicitly in
Table~\ref{table:boundary}. One can further test the foregoing result in
these sub-regimes by holding $\lao$, $\tilde\sigma_\perp$ and
$\sigpar$ fixed (as was done above) and decreasing
$|\sigH|/\tilde\sigma_\perp$ (or, equivalently, increasing
$\sigP/\tilde\sigma_\perp$) until (if $\lao$ is sufficiently small)
$\upo$ declines to a value $<1$ where, according to the hydrostatic
analysis, viable solutions cease to exist. 
Such calculations are presented in Fig.~\ref{fig:eta_sh}, where we plot
$|\sigH|/\tilde\sigma_\perp$ as a function of $\upo$ for representative
solutions in these two sub-regimes. It is seen that all viable solutions are
indeed restricted to the region $\upo > 1$, validating the above
prediction.

\subsubsection{Dependence on the field polarity}
\label{subsubsec:polarity}

As was noted in Section~I.2.1, the Ohm and Pedersen conductivities are
always positive, even under a global reversal of the field polarity,
since $\sigpar$ is not a function of the magnetic field strength and
$\sigP$ only contains magnetic terms that scale as $B^{2}$. The Hall
conductivity, however, has an overall linear dependence on $B \equiv
|\mathbf{B}|\, sgn\{B_z\}$ and can thus assume both positive and
negative values depending on the direction of the vertical field
component. The dependence of $\sigH$ on the magnetic field polarity was
shown in Paper~I to affect both the extent of the parameter ranges where
viable wind-driving disc solutions can exist in the Hall domain and the
properties of these solutions. We now briefly summarize these results,
which were obtained in the hydrostatic approximation assuming an
ion--electron plasma.

The range of values of the parameter $\beta \equiv 1/\beta_{\rm i0}$ for
viable solutions was found to be restricted by the following two
conditions (see Section~I.6):

\begin{enumerate}

\item \emph {Sub-Keplerian flow below the base of the wind}. This
implies that the normalized azimuthal velocity $w_\phi$ is $< 0$ within
the disc, which translates into ${\rm d}b_r/{\rm d}b_{\phi} < 0$ and
$\beta > - 2 \upo$ (equations~I.106--I.108).

\item \emph{Sub-Keplerian azimuthal velocity of the magnetic flux
surfaces}, or $w_{\rm Er0} = -r(\Omega_{\rm B0} - \Omega_{\rm K})/c_{\rm
s} > 0$, where $\Omega_{\rm B0} = -cE_{r0}/rB_{\rm0}$ is the angular
velocity of the flux surfaces at the disc mid-plane. This implies (see
equation~I.113) that either $\beta < -2 \upo$ or $\beta > - \upo/2$ in
the Hall domain. 
\end{enumerate}
The above two constraints together imply
\begin{equation}
\beta > -\upo/2 \qquad \mbox{in all the Hall sub-regimes} \,.
\label{eq:constr_all}
\end{equation}
By combining this inequality with the classification criterion
$\upo/|\beta| > 1/2$ that distinguishes the Hall sub-regimes~(i)
and~(iii) (see Section~\ref{subsec:illus}), one obtains 
\begin{equation} 
-\upo/2 < \beta < 2 \upo \qquad \mbox{Cases (i) and (iii)} \,.  
\label{eq:beta_constraint1}
\end{equation} 

\begin{figure}
   \centering
   \includegraphics[width=0.48\textwidth]{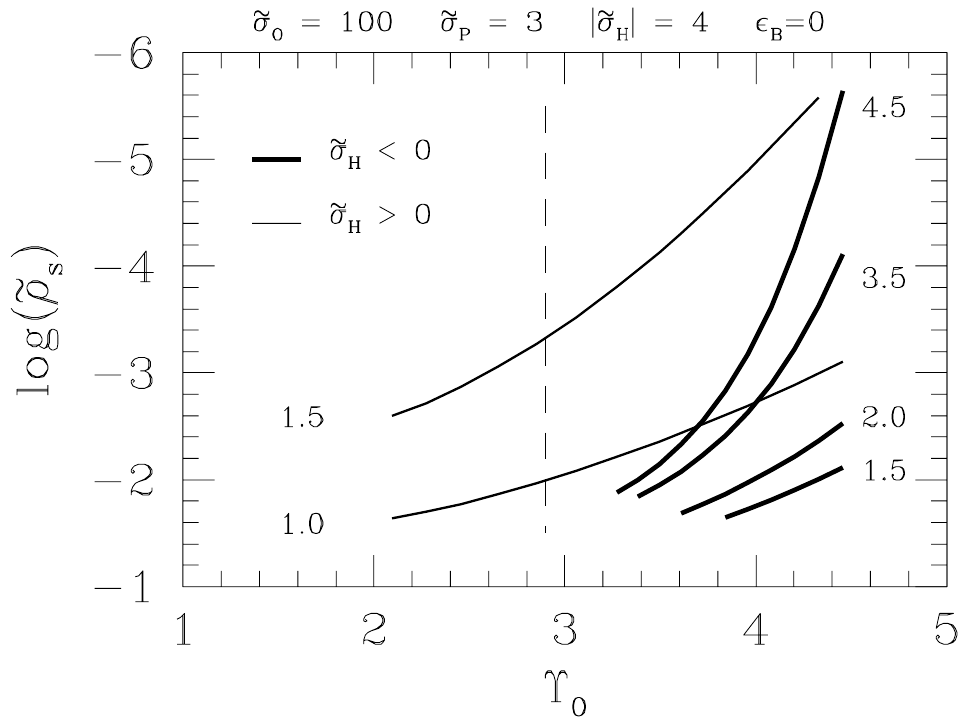} 
\caption{Normalized sonic-point density $\tilde \rho_{\rm s}$ as a
function of the parameter $\upo$ for solutions in the Hall
sub-regime~(i), illustrating the dependence of viable disc models on the
magnetic field polarity (i.e. the sign of $\sigH$). Curves for $\sigma_{\rm H} > 0$ are shown by
light lines and for $\sigma_{\rm H} < 0$ by heavy lines, and are labelled
by the normalized mid-plane inflow velocity $\epsilon$. 
(All the listed conductivities pertain to the mid-plane.)
The value of
$|\beta| \equiv 1/\betiabs$ is the same ($1.45$) for all the
solutions. The results are broadly consistent with the prediction of the
hydrostatic analysis (equation~\ref{eq:beta_constraint1}) that
there should be no viable solutions for $\sigH < 0$ when $\upo$
drops below $2|\beta|$ (i.e. to the left of the vertical dashed
line). 
}
\label{fig:5_26}
\end{figure}

Equation~(\ref{eq:beta_constraint1}) predicts that, although there could
be both positive- and negative-polarity solutions in the Hall
sub-regimes~(i) and~(iii), no viable solutions should exist in these
cases when $\beta$ decreases below $-\upo/2$. This prediction of the
hydrostatic analysis is examined in Fig.~\ref{fig:5_26}, which plots
$\tilde \rho_{\rm s}$, the normalized density at the sonic point, as a
function of the coupling parameter $\upo$ for Case-(i) solutions derived
using the indicated model parameters and corresponding to both positive
and negative values of $\beta \propto \sigH$ (see
equation~\ref{eq:betio}). Since the magnitudes of all the conductivity
tensor components are held constant, all the solutions are characterized
by the same absolute value of the ion Hall parameter ($\betiabs =
1/1.45$). The figure verifies that viable solutions cease to exist in
the region to the left of the vertical dashed line, which corresponds to
$|\beta| > \upo/2$, when $\beta < 0$.\footnote{Note that each of the
$\beta<0$ solution curves in Fig.~\ref{fig:5_26} terminates at a finite
value of $\upo$ that is lower the higher the value of $\epsilon$. To
understand this behaviour, note that, given the parameters that are held
fixed in this figure, $\upo$ scales as $a_0^2$. For a fixed value of
$a_0$, $\tilde \rho_{\rm s}$ increases as $\epsilon$ goes down (as seen also
in the top panel of Fig.~\ref{fig:5_10}). There is a maximum sonic-point
density $\tilde \rho_{\rm s}$ that can be attained for the chosen value
of $a_0$, corresponding to the maximum outflow rate for a consistent
solution, and this, in turn, determines the magnitude of $\epsilon$ for
the solution curve that terminates at the given value of $a_0$. A lower
value of $a_0$ corresponds to a higher lower bound on $\epsilon$ because
it implies a higher value of $|b_{\phi{\rm b}}|$ (see
equations~\ref{eq:B_rbB_phib}, \ref{eq:rphi_B} and~\ref{eq:upo}) and
hence a stronger torque on the disc and a correspondingly higher mass
accretion rate (reflected in the value of $\epsilon$).}

By similarly combining the inequality~(\ref{eq:constr_all}) with the
classification criterion $\upo/|\beta| < 1/2$ that distinguishes the
Hall sub-regimes~(ii) and~(iv), one obtains
\begin{equation}
\beta > 2 \upo \qquad \mbox{Cases (ii) and (iv)} \,, 
\label{eq:beta_constraint2}
\end{equation}
which implies that, for these sub-regimes, self-consistent solutions
exist only if the field has a positive polarity. 
We verified this prediction by adopting the values of
$a_{\rm 0}$, $\epsilon$ and $\tilde{\sigma}_\perp$ specified for
Cases~(ii) and~(iv) of Table~\ref{table:boundary} and changing the value
of $\sigH/\tilde{\sigma}_\perp$: we found that no negative-polarity
solutions could be obtained in these cases.

\begin{figure}%  figure placement: here, top, bottom, or page
   \centering
   \includegraphics[width=0.45\textwidth]{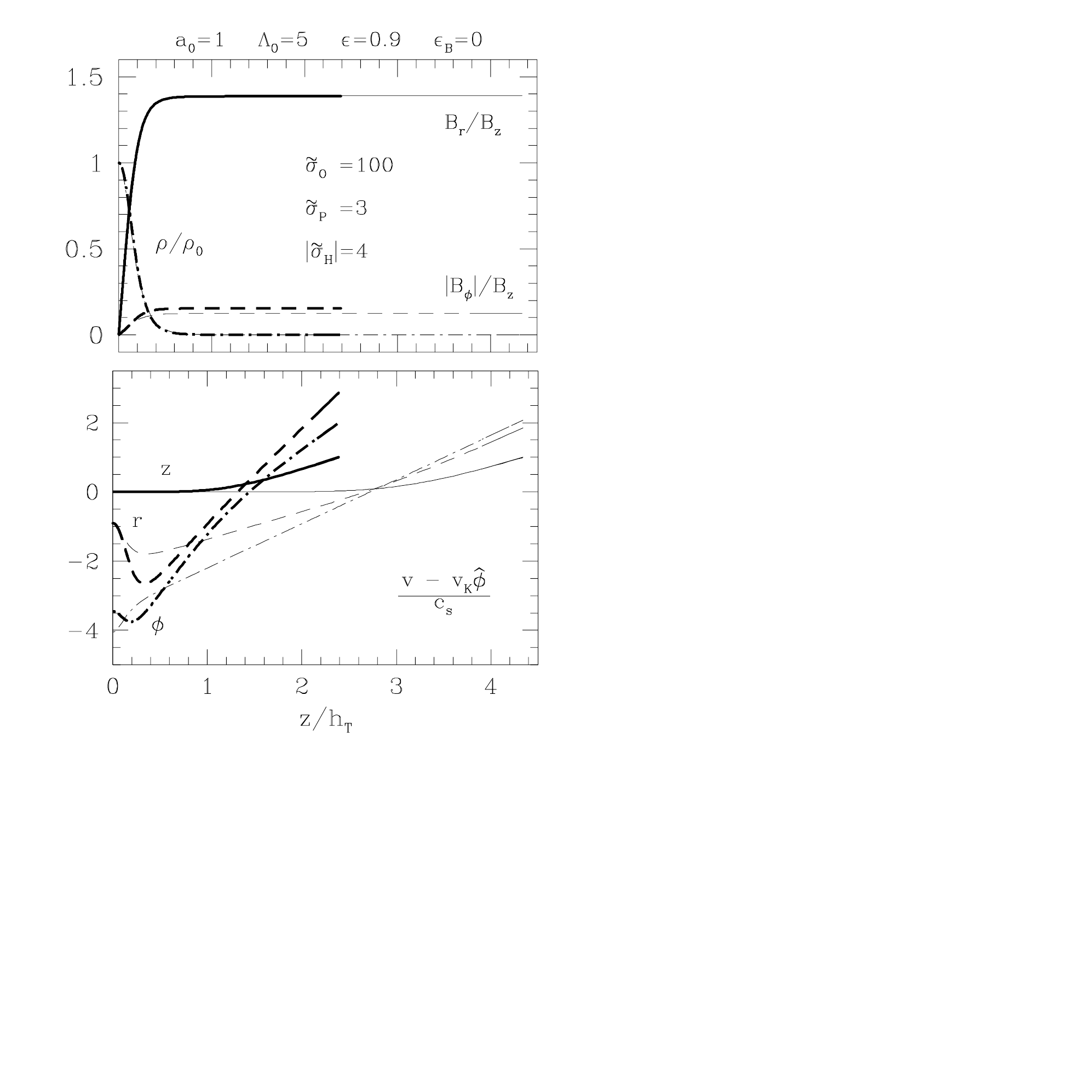} 
\caption{
Comparison of two disc solutions that differ only in the sign
of the Hall conductivity: $\sigH > 0$ for the solution shown by thin
lines, and $\sigH < 0$ for the solution depicted by thick lines. The
normalized upward mass flux $\rhot w_z$ is $9 \ee{-9}$ in the 
$\sigH > 0$ case and $5 \ee{-5}$ in the $\sigH < 0$ solution.
All the listed conductivities pertain to the mid-plane.
}
   \label{fig:5_2}
\end{figure}

   \begin{figure}%  figure placement: here, top, bottom, or page
   \centering
   \includegraphics[width=0.48\textwidth]{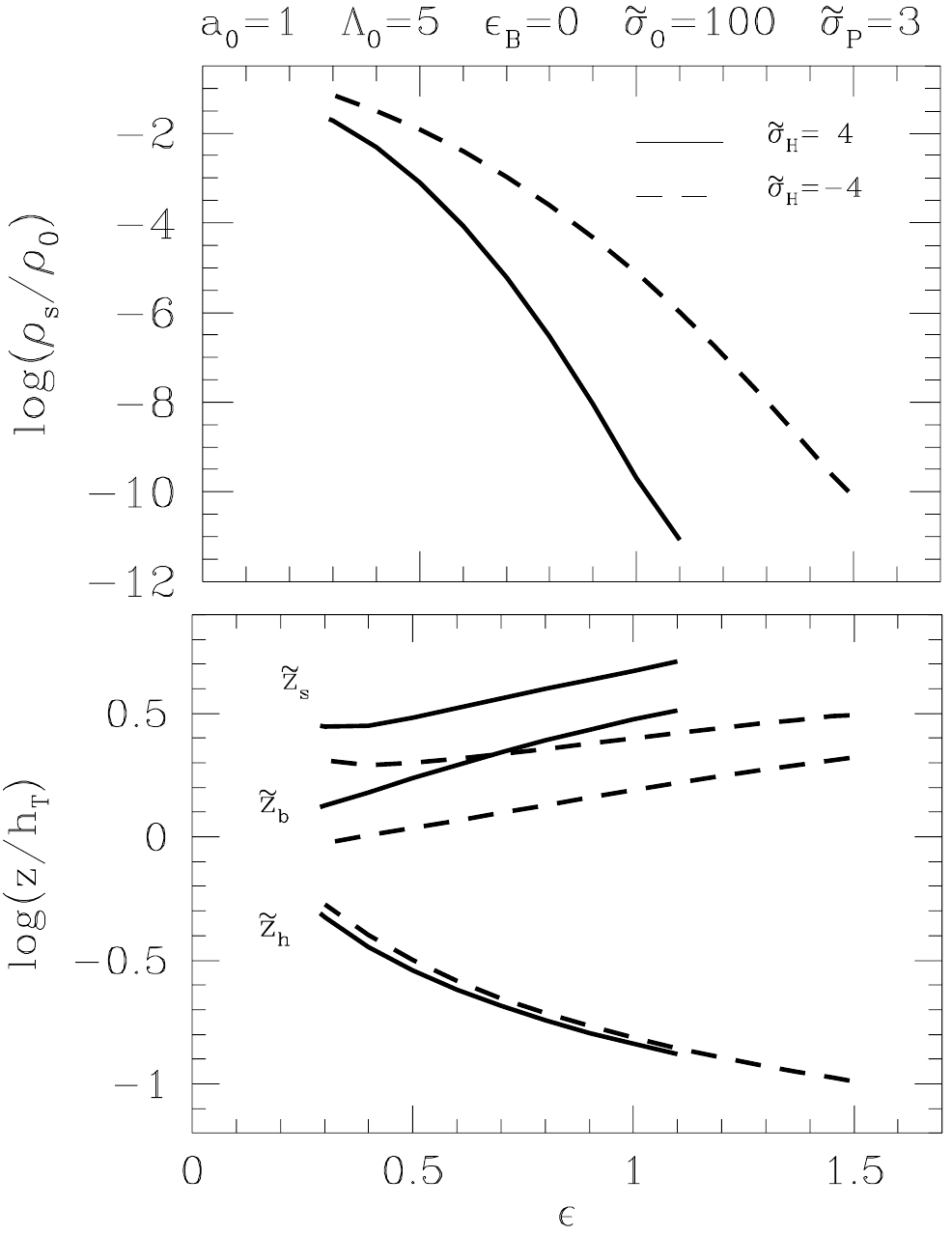} 
\caption{
Variation of the normalized sonic-point density (upper panel) and of the
characteristic heights $\tilde z_{\rm s}$, $\tilde z_{\rm b}$ and
$\tilde z_{\rm h} \equiv \tilde{h}$ (lower panel;
cf. Fig.~\ref{fig:5_10}) with the model parameter $\epsilon$ for the
same conductivity tensor components employed in Fig.~\ref{fig:5_2}. The
solid lines correspond to $\sigH > 0$ and the dashed lines to $\sigH <
0$.}
   \label{fig:5_3_1}
\end{figure}

Even when both positive and negative values of $\beta$ are allowed, as
is the case in sub-regimes~(i) and~(iii), changing the field polarity
modifies the properties of the solutions. This point, which was already
made in WK93, is illustrated in Figs.~\ref{fig:5_2} and~\ref{fig:5_3_1},
which compare disc solutions that differ only in the sign of
$\sigH$.\footnote{In discussing the differences between Hall-domain
solutions with opposite field polarities in Section~I.6 we explicitly
referred to sub-regime~(iii), whereas the solutions shown in
Figs.~\ref{fig:5_2} and~\ref{fig:5_3_1} correspond to sub-regime~(i). It
should, however, be clear from the present discussion that the behaviour
of solutions in these two sub-regimes is similar in this regard.}
The main difference between the two solutions in Fig.~\ref{fig:5_2} is
evidently the value of the sonic-point height $\zt_{\rm s}$
(corresponding to the termination point of the depicted solution), which
is larger ($\simeq 4.3$) in the positive-polarity case than in the
negative-polarity solution ($\simeq 2.4$). 
Fig.~\ref{fig:5_3_1} confirms
this trend: The wind-launching surface ($\zt_{\rm b}$) and the sonic surface
($\zt_{\rm s}$) are located higher above the mid-plane, and the mass
outflow rate (measured by $\tilde \rho_{\rm s}$) is correspondingly
lower, in the positive-polarity solutions. The
value of $\zt_{\rm h}=\tilde h$, which represents the scale on which the
density and transverse magnetic field components undergo their strongest
variation, is also different between the two sets of solutions (it is
larger for the negative-polarity model), although this difference is
modest in comparison with the change in $\zt_{\rm b}$ and $\zt_{\rm s}$,
and it goes in the opposite direction.

The above behaviour can be understood using the analytic expressions derived
in paper~I, and reproduced in Appendix \ref{sec:appA}. Employing the same reasoning as in our analysis of
Figs.~\ref{fig:10_1} and~\ref{fig:5_10} in Section~\ref{subsubsec:sigH},
we employ the height $\zt_{\rm b}$ of the base of the wind as a proxy for
$\zt_{\rm s}$ in this analysis. Using equations~(\ref{eq:ht})
and~(\ref{eq:zb}), we estimate that, to leading order, ${\tilde
h}_+/{\tilde h}_- \approx (2\upo - |\beta|)/(2\upo + |\beta|)$ and
$\zt_{\rm b +}/\zt_{\rm b -} \approx (\upo^2 + 5\upo |\beta|/2 +
\beta^2)/(\upo^2 - 5\upo |\beta|/2 + \beta^2)$, where the subscripts `+'
and `-' refer to the positive- and negative-polarity cases,
respectively. Using also equations~(\ref{eq:upo}) and~(\ref{eq:betio}),
we infer that, for the parameters adopted in Figs.~\ref{fig:5_2}
and~\ref{fig:5_3_1}, ${\tilde h}_+/{\tilde h}_- \approx 0.85$ and
$\zt_{\rm b +}/\zt_{\rm b -} \approx 2.28$. These estimates are entirely
consistent with the numerical results (for a fixed value of $\epsilon$)
shown in the two figures. Furthermore, from
equation~(\ref{eq:B_rbB_phib}) we confirm that $|b_{r{\rm
b}}/b_{\phi{\rm b}}|$ is $\gg 1$ and hence (using
equation~\ref{eq:rphi_B}) that $b_{r{\rm b}} \approx \sqrt{2}/a_0$,
independently of the sign of $\beta$, reproducing the actual behaviour of
the solutions in Fig.~\ref{fig:5_2}. Setting $b_{r{\rm b+}}\approx
b_{r{\rm b-}}$, we then infer from equations~(\ref{eq:B_rbB_phib})
and~(\ref{eq:ht}) that $b_{\phi{\rm b}+}/b_{\phi{\rm b}-} \approx
{\tilde h}_+/{\tilde h}_-$ ($\approx 0.85$ for the adopted
parameters). This conclusion, too, is consistent with the result
exhibited in Fig.~\ref{fig:5_2}.

Although the specific angular momentum at the base of the flow is mostly
magnetic (corresponding to the wind model parameter $\lambda$ being $\gg
1$), the fact that $b_{\phi{\rm b}+}/b_{\phi{\rm b}-} < 1$ does not
imply that the value of $\lambda$ is larger in the negative-polarity
case. In fact, the converse is true, as can be seen using
equation~(\ref{eq:ele}), from which it follows that $\lambda_+/\lambda_-
\approx (b_{\phi{\rm b}+}/b_{\phi{\rm b}-})(\rhot_{\rm s -}/\rhot_{\rm s
+})$. Given that $\rhot_{\rm s -}/\rhot_{\rm s+}$ is typically $\gg
b_{\phi{\rm b}-}/b_{\phi{\rm b}+}$ (as the solutions in
Fig.~\ref{fig:5_2} demonstrate), we find that $\lambda_+/\lambda_- 
\gg 1$. Physically, the magnetic torque acting on the disc ($\propto B_z
B_\phi$) and correspondingly the mass accretion rate and the rate of
inward angular momentum advection are slightly larger in the
negative-polarity case. However, the mass outflow rate is significantly
larger in this case and therefore the \emph{specific} angular momentum
(the angular momentum per unit mass) has to be much smaller in order for
the rate of inward (radial) and outward (vertical) angular momentum
transport to balance each other. An alternative way of arriving at this
conclusion is to consider the wind solutions that are self-consistently
matched to these disc solutions. The matched wind solutions lie on a
$\xi_{\rm b}{'}=\;$const curve in the $\kappa-\lambda$ wind parameter
space (see equation~\ref{eq:incl}), and along such a curve \emph{higher}
values of $\kappa \propto \rhot_{\rm s}$ (see equation~\ref{eq:kappa})
correspond to \emph{lower} values of $\lambda$ (see Fig.~2 in BP82).

\begin{figure}%  figure placement: here, top, bottom, or page
   \centering
   \includegraphics[width=0.45\textwidth]{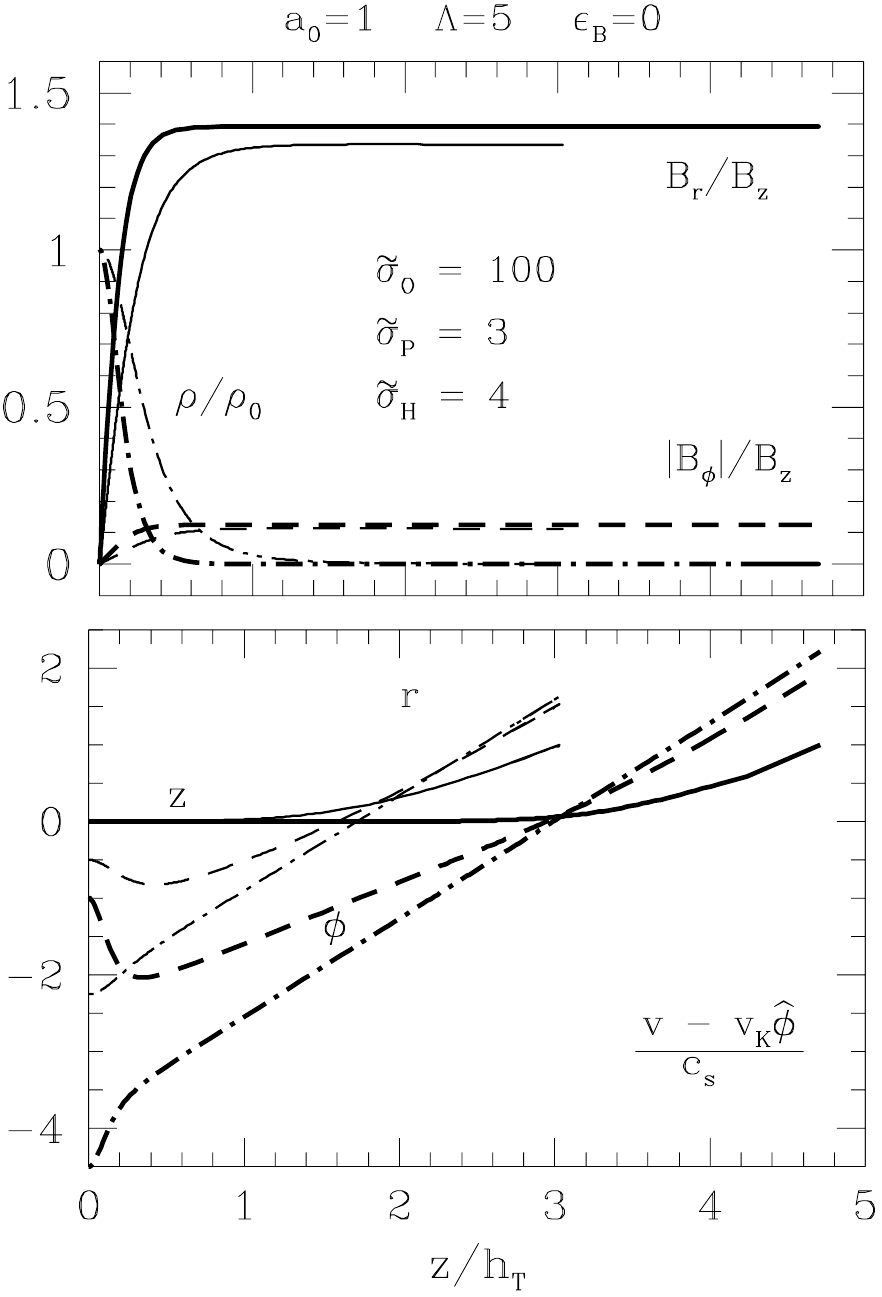} 
\caption{Vertical structure of a wind-driving disc solution for two
different values of the radial velocity parameter: $\epsilon =
1$ (thick lines) and $\epsilon=0.5$ (thin lines). The other model
parameters are listed in the figure 
(with all the conductivity values pertaining to the mid-plane).}
   \label{fig:5_6_3}
\end{figure}

\begin{figure}
   \centering
   \includegraphics[width=0.45\textwidth]{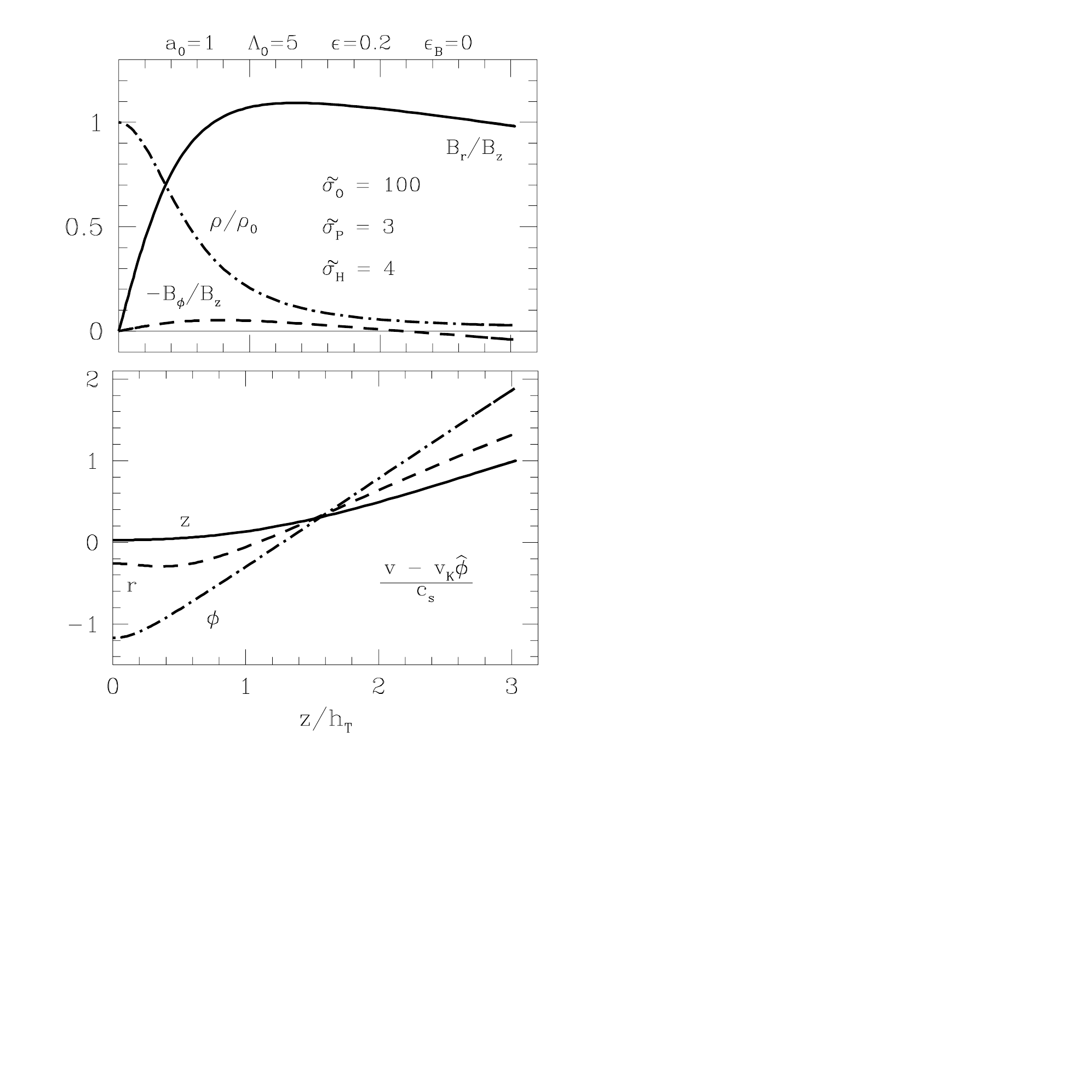} 
   \caption{Same as the left-hand side of Fig.~\ref{fig:i-ii}, but for
$\epsilon = 0.2$. In this case $\upo = 9.1$ (see
Table~\ref{table:boundary}), so the inequality
$\epsilon \upo > 2$ (the third constraint on viable solutions in the
Hall sub-regime~(i); see the first row in Table~\ref{table:boundary}) is
not satisfied. As discussed in the text, the resulting disc solution is
unphysical.}
\label{fig:5_9} 
\end{figure}

\subsubsection{Dependence on the radial velocity parameter $\epsilon$}
\label{ref:subsubsec:epsilon} 

To illustrate the dependence on the parameter $\epsilon$ (the normalized
mid-plane inflow velocity), we plot in Fig.~\ref{fig:5_6_3} two
solutions in the Hall sub-regime~(i) that differ only by the value of
this parameter. It is seen that, as $\epsilon$ decreases from 1.0 (thick
lines) to 0.5 (thin lines), the scaleheight $\tilde h$ (the scale on
which the density and transverse magnetic field components vary most
strongly) increases and $\zt_{\rm s}$ (the vertical extent of the
displayed solution, corresponding to the location of the sonic surface)
decreases. These trends are also evident 
in the solutions depicted in 
Figs.~\ref{fig:5_10} and ~\ref{fig:5_3_1}. As
we did in Sections~\ref{subsubsec:sigH} and~\ref{subsubsec:polarity}, we
use $\zt_{\rm b}$ as a proxy for $\zt_{\rm s}$ and employ the
hydrostatic-approximation equations~(\ref{eq:ht}) and~(\ref{eq:zb}) to
analyze this behaviour. These two expressions show that, with all the
other parameters remaining unchanged, ${\tilde h} \propto 1/\epsilon$
and $\zt_{\rm b} \propto \epsilon$, which can directly account for the
exhibited trends. In both cases, the dependence on $\epsilon$ can be
traced to the scaling $\jr \propto \wr$ implied by the angular momentum
conservation relation (equation~\ref{eq:phi_motiond}) in the hydrostatic
limit (see Section~I.4.3).

The hydrostatic analysis presented in Paper~I also leads to a lower limit on
the value of $\epsilon$, derived from the requirement that the base
of the wind be located above a density scaleheight, i.e. $\zt_{\rm b} >
{\tilde h}$. This implies $\epsilon \upo > (18)^{1/4} \approx 2$ (see
Table~\ref{table:constraints1}) and represents the third constraint
listed in Table~\ref{table:constraints}. To verify the applicability of
this condition, we constructed a solution using the same parameters as
in the illustrative Case-(i) solution depicted in Fig.~\ref{fig:i-ii}
except that we chose a smaller value of $\epsilon$ so that the above
constraint is no longer satisfied. The result, shown in
Fig.~\ref{fig:5_9}, confirms that, when the above inequality is violated,
the solution is no longer physically viable. This is evidenced
by the downward turn of the $|b_\phi|$ and $b_r$ curves above a certain
height. A similar solution was presented in Fig.~6 of WK93, who
explained the origin of this behaviour by noting that, when $\epsilon$
decreases to a sufficiently low value, $\zt_{\rm s}$ becomes so small
and (correspondingly) $\rhot_{\rm s}$ so large that the upward mass flux
carries more angular momentum than is brought in by the accretion
flow. Consequently, the gradient of $b_\phi$ changes sign (with that of
$b_r$ following suit) as the magnetic field starts depositing angular
momentum back into the flow even before the nominal top
of the disc is reached. Such a configuration is likely unstable
\citep[e.g.][]{CS94}.

\subsection{Matched disc--wind solutions}
\label{subsec:global}

  \begin{figure}%  figure placement: here, top, bottom, or page
   \centering
   \includegraphics[width=0.48\textwidth]{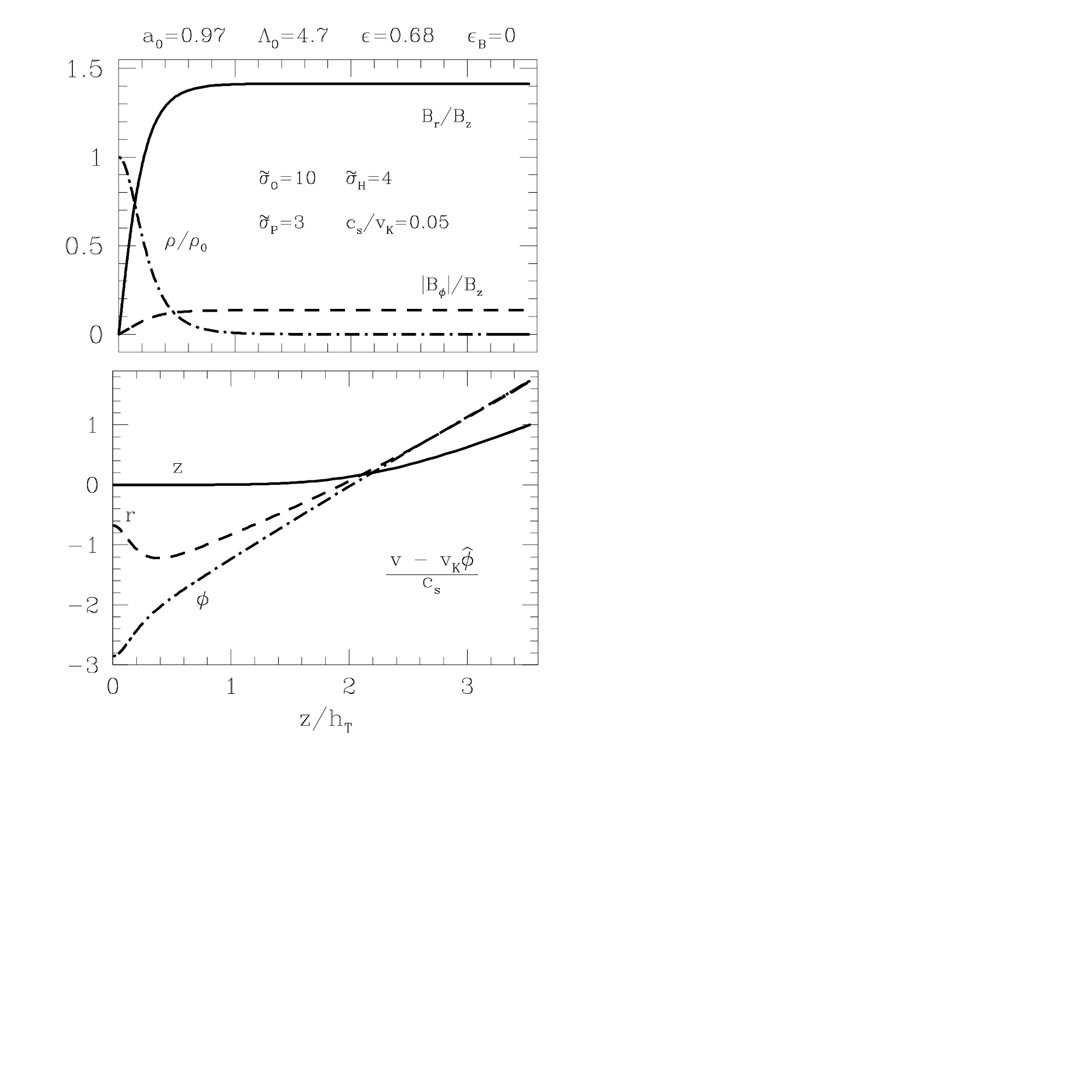} 
\caption{
Illustrative disc--wind solution obtained by matching a radially
localized disc solution to a global (self-similar) wind solution using
the procedure outlined in Section~\ref{subsec:globalsol}. The disc model
parameters are shown in the figure 
(with all the conductivity values pertaining to the mid-plane)
whereas those of the matched wind solution are $\kappa = 7.9 \ee{-4}$,
$\lambda = 174$ and $\xi'_{\rm b} = 1.42$. This solution satisfies the
constraints specified in Table~\ref{table:constraints} for the Hall
conductivity sub-regime (iii): $0.1 \lesssim 0.97 \lesssim 2 \lesssim
6.4 \lesssim 10$.}
   \label{fig:5_18}
\end{figure}

 \begin{figure}%  figure placement: here, top, bottom, or page
   \centering
   \includegraphics[width=0.48\textwidth]{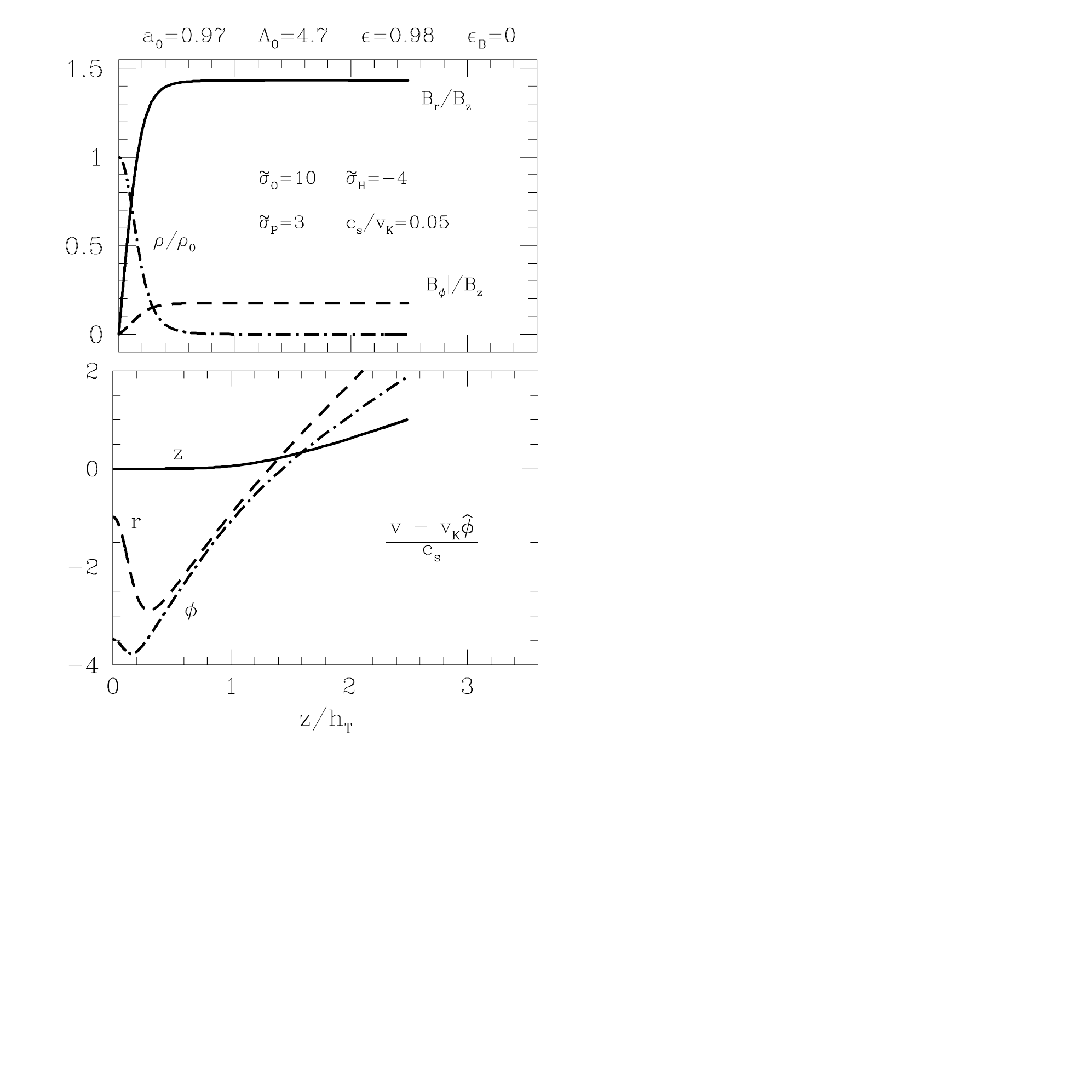} 
\caption{
Same as Fig.~\ref{fig:5_18}, but with $\sigH$ having the opposite
sign. All the other disc model parameters except for $\epsilon$, which
is determined by imposing the Alfv\'en regularity condition on the wind
solution, are the same as in Fig.~\ref{fig:5_18}. In this case the
parameters of the matched wind solution are $\kappa = 1.2 \ee{-3}$,
$\lambda = 144$ and $\xi'_{\rm b} = 1.43$. This solution, too, satisfies the
hydrostatic-analysis constraints for the Hall conductivity sub-regime
(iii): $0.1 \lesssim 0.97 \lesssim 2 \lesssim 9.2 \lesssim 10$.}
   \label{fig:5_19}
\end{figure}

The solutions presented in Sections~\ref{subsec:illus}
and~\ref{subsec:test} have involved the disc model alone. This was done
in order to simplify the derivations and was an adequate approach given
that we were primarily interested in studying the constraints on the
disc model parameters. However, to justify this treatment, it is
necessary to demonstrate that matched disc--wind solutions can be
obtained for similar disc parameter combinations. This is done in this
subsection, where we present two examples of self-consistently matched
(local) disc and (global) wind solutions. These solutions correspond to
the same values of the parameters $\lao$ and $a_0$ and of the absolute
values of the conductivity tensor components but have opposite signs of
$\sigH$: they were obtained using the procedure described in
Section~\ref{subsec:globalsol} and are shown in Figs.~\ref{fig:5_18} (the
positive-polarity case) and~\ref{fig:5_19} (the negative-polarity case).
As is evident from equations~(\ref{eq:kappa}) and~(\ref{eq:ele}),
combining the two solutions requires a specification of the parameter
$c_{\rm s}/\vk = h_{\rm T}/r$: the above examples were computed for
$c_{\rm s}/\vk = 0.05$.

Our matched solutions correspond to the Hall sub-regime~(iii) and are seen
to be very similar to the analogous solutions we obtained by considering
the disc alone (right panels of Fig.~\ref{fig:i-ii}).
The qualitative differences
between the $\sigH > 0$ and $\sigH < 0$ solutions (essentially a higher
value of $\zt_{\rm s}$ and lower values of $\tilde h$ and of $|b_\phi|$
in the positive-polarity case) are also the same as those found in the
`windless' solution (Figs.~\ref{fig:5_2} and~\ref{fig:5_3_1}). This is
as expected, since the matching to the wind solution does not affect the
physical requirements on a viable disc solution; rather, what it does is
fix -- through the imposition of the Alfv\'en regularity condition on
the wind solution -- the value of one of the disc model parameters that
is chosen arbitrarily when the wind solution is not taken into
account. In our treatment (see Section~\ref{subsec:globalsol}), the disc
parameter that is determined in this way is $\epsilon$: it is equal to
0.68 in the $\sigH > 0$ solution shown in Fig.~\ref{fig:5_18} and to
0.98 in the $\sigH < 0$ solution shown in Fig.~\ref{fig:5_19}. In an
even more self-consistent solution, the value of the parameter
$\epsilon_{\rm B}$ would also be determined by taking into account the
conditions outside the disc -- in this case the magnetic flux
distribution along the disc surface (see Section~I.3.1.3) -- rather than
arbitrarily setting it equal to zero as has been done in this paper and
in the analytic derivations of Paper~I.\footnote{A self-consistent
treatment along these lines has, in fact, already been implemented in
the fully global self-similar disc/wind model constructed by
\citet{Tei10}.}

\section{Conclusion}
\label{sec:conclude}

In this paper we continue our study of the viability and properties of
weakly ionized protostellar accretion discs that transport their excess
angular momentum vertically through the surfaces by means of
centrifugally driven winds. In view of the suggestive evidence that this
situation is realized in at least some protostellar systems
\citep[e.g.][]{Ray07}, and yet mindful of the fact that the total radial
extent of such wind-driving disc regions is still unknown, we consider
radially localized disc configurations (which, however, are joined to a
global wind model). We employ the formulation devised by WK93 for
modelling steady, geometrically thin, vertically isothermal and nearly
Keplerian discs in which magnetic diffusivity counters the shearing and
advection of the magnetic field. WK93 assumed that the charge carriers
were singly-charged ions and electrons and focussed on the ambipolar
diffusivity regime. In Paper~I we extended this model to the Hall and
Ohm diffusivity regimes using the conductivity-tensor formalism,
although we reverted to using the multifluid approach employed by WK93
for deriving parameter constraints on physically viable solutions in
these regimes. This derivation generalized the corresponding results of
WK93 for the ambipolar diffusivity regime and was similarly carried out in
the context of the hydrostatic approximation, in which the vertical
velocity component is neglected inside the disc (which results in
several of the differential equations for the disc structure simplifying
to algebraic relations).

The hydrostatic analysis of paper~I indicated that viable wind-driving
disc solutions correspond to four parameter sub-regimes in the Hall
diffusivity domain and three in the Ohm domain. It also led to
analytic estimates of the magnetically reduced (due to magnetic
pressure-gradient `squeezing') density scale height and of the location
of the disc's surface (where the inflow turns into an outflow), as well
as of other pertinent quantities. These results are summarized in
Appendix~\ref{sec:appA}. In this paper we test these predictions by
constructing exact solutions of the disc equations. We concentrate on
the Hall regime in view of its expected importance in the inner regions
of real systems; we do not consider the low-ionization Ohm regime in
this paper given that it may have limited relevance to wind-driving
protostellar discs. We characterize the solutions in terms of the
conductivity-tensor components (i.e. the Pedersen, Hall and Ohm
conductivities). However, to facilitate the comparison with the analytic
results of Paper~I, which were derived in the framework of the multifluid
formulation, we assume that the ratios of these terms are constant with
height in the disc and, more specifically, that they scale with the
density and field amplitude as $\rho/B^2$, which implies that the
matter--field coupling parameter (the Elsasser number $\Lambda$) is also
constant with height. We require the derived solutions to cross the
sonic critical surface but we do not continue the integration past that
surface; this is sufficient for the comparison with the analytic results
and greatly simplifies the calculations. However, as recapitulated below,
we also demonstrate that these solutions can be matched to wind
solutions that extend to large distances (and, in particular, cross the
Alfv\'en critical surface). Our findings can be summarized as follows.

\begin{enumerate}

\item Our numerical solutions are in broad agreement with the
parameter constraints obtained under the hydrostatic approximation for
the four Hall parameter sub-regimes.  In the regions of
parameter space that are excluded by the above constraints, wind-driving
disc solutions cannot be obtained or are unphysical.

\item All viable solutions satisfy the constraint $\upo \gtrsim 1$ (see
equation~\ref{eq:eta}). Physically, this condition expresses the
requirement that the mid-plane neutral--ion momentum exchange time be
shorter than the disc orbital time. As discussed in Paper I, this
requirement is predicted to apply in each of the diffusivity regimes and
is evidently a fundamental constraint on disc solutions of the type
considered here.

\item For the same values of $a_{\rm 0}$, $\lao$ and $\epsilon$ (the
normalized mid-plane magnetic field amplitude, matter--field coupling
strength and radial speed, respectively), increasing the relative
contribution of the Hall
conductivity 
(the ratio $|\sigH|/\tilde \sigma_\perp$) results in a smaller (magnetically
reduced) density scaleheight $\tilde{h}$) and in larger heights of the disc
and sonic surfaces ($\zt_{\rm b}$ and $\zt_{\rm s}$,
respectively). A higher value of $\zt_{\rm s}$ in turn implies a lower
density at the sonic surface ($\rhot_{\rm s}$) and hence a lower mass
outflow rate.

\item The magnetic field polarity affects both the properties of the
solutions and the extent of the parameter regimes where viable solutions
exist when the Hall current is dynamically important, reflecting the
dependence of the Hall conductivity on the sign of $B_z$. Specifically,
we confirmed the following dependence of the solutions on the field
polarity (with a positive polarity corresponding to $B_z$ being
parallel to the disc rotation vector):

\begin{enumerate}

\item {\it Hall sub-regimes (i) and (iii)}. The parameter $\beta$, 
which is equal to the inverse of the mid-plane ion Hall parameter
(equation~\ref{eq:bebiq}) and thus scales inversely with the {\it
signed}\/ magnetic field amplitude $B$, is predicted to lie in the range
$-\upo/2<\beta<2\upo$. We verified that no solutions exist for $\beta <
-\upo/2$ and that positive- and negative-polarity solutions have
distinct properties. In particular, when the sign of $\sigH$ is changed
from $<0$ to $>0$ and all the other parameter values remain the same,
$\tilde{z}_{\rm b}$ and $\tilde{z}_{\rm s}$ are increased and the wind
outflow rate correspondingly goes down.

\item {\it Hall sub-regimes (ii) and (iv)}. These parameter regimes are
characterized by $\beta > 2 \upo$. We verified that only
positive-polarity solutions can be obtained in this case.

\end{enumerate}

\item Decreasing the inward radial speed ($\epsilon$) \emph{increases} the
scale over which the fluid variables vary most strongly (the
magnetically reduced density scaleheight $\tilde{h}$) but
\emph{decreases} both $\zt_{\rm b}$ and $\zt_{\rm s}$.
Furthermore, when the lower limit on the value of $\epsilon$,
obtained from the requirement that $\zt_{\rm b} > \tilde{h}$ (the
third inequality in Table~\ref{table:constraints}) is violated, the
computed solutions are found to be unphysical (in that $b_r$ and
$|b_\phi|$ start to decrease with $\zt$ above a certain height).

\end{enumerate}

We also detail the procedure for obtaining global (radially
self-similar) `cold' wind solutions following the methodology introduced
by BP82. We compute solutions of this type for a large range of values
of the wind model parameters $\kappa$, $\lambda$ and $\xi'_{\rm b}$
(the normalized mass-to-flux ratio, specific angular momentum and
field-line inclination at the disc's surface, respectively). Tables of
these solutions are available on the VizieR data base of astronomical
catalogues (http://cdsarc.u-strasbg.fr/). As our radially localized and
geometrically thin model cannot be used to follow the propagation of the
outflow far from the disc, we match our disc solution to a BP82-type
wind solution by adjusting one of the disc model parameters ($\epsilon$)
and iterating on the disc and wind calculations until the full solution
converges. We present illustrative solutions of this type that
demonstrate that matched disk/wind configurations can be obtained for
parameter values that are very similar to those of the merely transonic
solutions employed in our parameter-space analysis.

The accretion process in protostellar discs may involve a variety of
angular-momentum transport mechanisms, including, in particular, radial
transport by gravitational torques and by MRI-induced turbulence. In
this paper we consider only vertical transport by centrifugally driven
winds in an attempt to model a radially localized disc region where this
mechanism may dominate. (Note, however, that both vertical transport and
radial transport -- notably MRI-induced turbulence -- could in principle
operate at the same disc radius; see \citealt*{SKW07}.) As discussed in
Paper~I, the large-scale, ordered magnetic field envisioned in this
scenario could be either interstellar field advected by the accretion
flow or dynamo-generated field produced in either the star or the
disc. In view of the strong evidence for strong outflows from the inner
regions of protostellar discs, we also neglect alternative modes of
angular momentum transport that could be mediated by such a field,
including magnetic braking, `failed' winds and non-steady phenomena. Our
treatment has been deliberately simplified to facilitate comparison with
the analytic results of Paper~I; in particular, we assume that the
matter is everywhere well coupled to the field (i.e. $\Lambda > 1$) and
that the same conductivity regime applies between the mid-plane and the
disc's surface. In reality, the disc may be weakly coupled between the
midplane and some finite height and its diffusivity properties are
expected to change with $\zt$ \citep[e.g.][]{SW05}. Our approximation
should be adequate for representing the behaviour of the dominant
diffusivity regime in the well-coupled region of the disc. However, if a
significant fraction of the local column density is magnetically weakly
coupled then the vertically averaged properties of the disc (such as the
inflow speed) could be significantly modified \citep[see][]{Li96,War97}.

In conclusion, the results presented in this paper confirm the validity
of the parameter constraints derived in Paper I for physically viable
configurations of Hall diffusivity-dominated protostellar discs in which
centrifugally driven winds dominate the local angular momentum
transport. They also demonstrate that the algebraic expressions derived
on the basis of the hydrostatic approximation correctly identify the
generic properties of such discs and are useful for clarifying the
behaviour of the full numerical solutions. More generally, the
theoretical framework developed in WK93, Paper~I and the present work
can be used to study discs of this type also in other diffusivity
regimes and in other astrophysical environments. In particular, it can
help interpret observations of such systems by relating the properties
of the outflow to those of the underlying disc
\citep[e.g.][]{Kon10}. 
It may also be useful for guiding non-ideal-MHD numerical simulations of
wind-driving discs.

\section*{Acknowledgments}
We thank Matthew Kunz for useful discussions and suggestions, and are grateful for the hospitality provided by the Isaac Newton Institute for Mathematical Sciences at Cambridge University, where some of this work was conducted. This research was supported in part by NASA Theoretical Astrophysics
Programme grant NNG04G178G (AK and RS), by NSF grant AST-0908184 (AK)
and by the Australian Research Council grants DP0344961, DP0881066 (RS
and MW) and DP0342844 (RS).

\appendix
\section[]{Parameter constraints in the Hall limit}
\label{sec:appA}

The parameter constraints on viable wind-driving disc solutions were
obtained in Paper~I by applying the hydrostatic approximation to a
radially localized disc model under the assumption that the charged
component of the weakly ionized disc material consisted of two particle
species: positive `ions' (subscript `i') and negative `electrons'
(subscript `e'), each singly charged. Instead of employing two
independent ratios of the conductivity-tensor components as model
parameters, as is done in this paper, we used the ion and electron Hall
parameters, i.e. the ratios of the gyrofrequency and the collision
frequency of these two species with the neutrals, which are given by
\begin{equation} 
\beta_{\rm e} = \frac{eB}{m_{\rm i} c} \, \frac{1}{\gamma_{\rm i}
\rho}\; , \quad\quad  \beta_{\rm i}= \frac{eB}{m_{\rm e} c} \,
\frac{1}{\gamma_{\rm e} \rho} \equiv q\, \beta_{\rm e}\, ,
\label{eq:bebiq}
\end{equation}
where $c$ is the speed of light, $m$ is the particle's mass, $e$ is the
unit electric charge and the total mass density $\rho$ is used to
approximate the neutral mass density.  The collisional coupling
coefficient $\gamma_j$ is equal to $\langle \sigma v \rangle_j/ (m_j +
m) $, where $m$ is the mean mass of the neutral particles and $\langle
\sigma v \rangle_j$ is the rate coefficient of momentum exchange of
species $j$ with the neutrals. The product $\gamma_{\rm i} \rho$
($\gamma_{\rm e} \rho$) in equation~(\ref{eq:bebiq}) therefore
represents the momentum-exchange collision frequency of the ions
(electrons) with the neutrals. In the above definitions, $B \equiv
|\mathbf{B}|\, sgn\{B_z\}$, so the sign of the Hall parameter is
sensitive to the magnetic field polarity. It is further assumed that the
ions are `heavy' and the electrons are `light', so that $q \ll
1$.\footnote{In a real disc containing two charged species, the value of
$q$ is fixed by the physical properties of the charge carriers (see
equations~I.7--I.9). This, in turn, constrains the values that the
conductivity ratios $\sigH/{\tilde \sigma_\perp}$ and
$\tilde\sigma_\perp/\sigpar$ can take (see equations~\ref{eq:betio}
and~\ref{eq:beteo}). We do not incorporate this constraint into the
parameter-space analysis in Section~\ref{sec:results} so as not to
unduly complicate the discussion; however, we have verified that the
`effective' value of $q$, obtained from the ratio of
equation~\ref{eq:betio} to equation~\ref{eq:beteo}, remains $\ll 1$ for
all the solutions that we present.}

Another key parameter employed in Paper~I (see also WK93) is
\begin{equation}
\Upsilon \equiv \frac{\gamma_{\rm i} \rho_{\rm i}}{\Omega_{\rm K}}
\label{eq:eta}
\end{equation}
(where $\rho_{\rm i}$ is the ion mass density), the ratio of the
Keplerian rotation time to the neutral--ion momentum exchange time. In
the ambipolar diffusivity limit ($\sigma_{\rm O} \gg \sigma_{\rm P} \gg
|\sigma_{\rm H}|$ or, equivalently, $|\beta_{\rm i}| \gg 1$) the
Elsasser number $\Lambda$ (Section~\ref{subsec:param}) reduces to
$\Upsilon$, whereas in the Hall diffusivity limit ($\sigma_{\rm P} \ll
|\sigma_{\rm H}| \ll \sigma_{\rm O}$ or, equivalently, $|\beta_{\rm i}|
\ll 1 \ll |\beta_{\rm e}|$) $\Lambda = \Upsilon |\beta_{\rm i}|$.

\setlength{\cellspacetoplimit}{1mm}
\setlength{\cellspacebottomlimit}{1mm}
\begin{table*}
\caption{Parameter constraints for wind-driving disc solutions in the
limit where the Hall diffusivity dominates and assuming $\epsilon_{\rm
B} = 0$. Four distinct cases can be identified, depending on how the
values of $s_0 = \beta_{\rm e 0} \beta_{\rm i 0}$ and of $2\Lambda_0 =
2\Upsilon_0 |\beta_{\rm i 0}|$ compare with 1. The first inequality
expresses the requirement that the disc remain sub-Keplerian below the
wind zone ($\tilde{z} < \tilde{z}_{\rm b}$), the second is the wind
launching condition (the requirement that the magnetic field lines be
sufficiently inclined to the vertical for centrifugal acceleration to
occur), the third ensures that the base of the wind is located above the
(magnetically reduced) density scaleheight and the fourth specifies
that the rate of Joule heating at the midplane should not exceed the
rate of release of gravitational potential energy there.}
\label{table:constraints} 
\begin{center}
\begin{tabular}{ScScScScScScScScScScScSc} 
\hline 
{\normalsize Case} & \multicolumn{2}{c}{{\normalsize Limits}} &
\multicolumn{9}{c}{{\normalsize Parameter Constraints -- Hall Limit }}\\ 
& \ \ \ $s_0=\beta_{\rm e 0} \beta_{\rm i 0}$ & \ \ \ $\Lambda_0=\Upsilon_0 |\beta_{\rm i
0}|$ & \multicolumn{9}{c}{{\small (multi-fluid formulation)}}\\ 
\hline 
(i) & $> 1$ & $> 1/2$ & \ \ \ \ \ $(2\Upsilon_0)^{-1/2}$ & $\lesssim$ & $a_0$ &
$\lesssim$ & $2$ & $\lesssim$ & $\epsilon\Upsilon_0$ & $\lesssim$ & $\vk/2
c_{\rm s}$ \\ 

(ii) & $ > 1$ & $< 1/2$ & \ \ \ \ \ $\betio^{1/2}$ & $ \lesssim$ &
$a_0$ & $\lesssim$ & $2 (\Upsilon_0 \betio)^{1/2}$ & $\lesssim$ & $\epsilon/
2\betio$ & $\lesssim$ & $\Upsilon_0 \betio \vk/ c_{\rm s}$ \\ 

(iii) & $< 1$ & $>
1/2$ & \ \ \ \ \ $(2\Upsilon_0)^{-1/2}$ & $\lesssim$ & $a_0$ & $\lesssim$ & $2$ &
$\lesssim$ & $\epsilon \Upsilon_0 \beta_{\rm e0} \beta_{\rm i0}$ & $\lesssim$
& $\vk/2 c_{\rm s}$ \\ 

(iv) & $< 1$ & $< 1/2$ & \ \ \ \ \ $\betio^{1/2}$ & $\lesssim$ & $a_0$ & $\lesssim$
& $2(\Upsilon_0 \betio)^{1/2}$ & $\lesssim$ & $\epsilon \beta_{\rm e0}/2$ &
$\lesssim$ & $\Upsilon_0 \betio \vk/ c_{\rm s}$ \\ 
\hline 
\end{tabular} 
\end{center} 
\end{table*}

The parameter constraints for the four Hall sub-regimes are presented in
Table~\ref{table:constraints}, which reproduces Table~I.1. The physical
origin of the imposed constraints is summarized in the caption of this
table. The key predicted properties of the solutions in each sub-regime
are listed in Table~\ref{table:constraints1}, which
reproduces Table~I.2. We also reproduce below some of the expressions
used in the derivation of these results that are relevant to
the analysis presented in this paper.

\setlength{\cellspacetoplimit}{1mm}
\setlength{\cellspacebottomlimit}{1mm}
\begin{table*}
\caption{Key properties of viable disc solutions in the Hall
regime. Listed, in order, are the midplane values of $|db_r/db_\phi|$,
the magnetically reduced scaleheight in units of the tidal
scaleheight ($\tilde h \equiv h /h_{\rm T}$), the similarly normalized
vertical location of the base of the wind $\tilde{z}_{\rm b}$ in units
of $\tilde{h}$ and the normalized Joule dissipation rate $\j \bdot
{\bmath e}^\prime$ at the midplane.}
\label{table:constraints1}
\begin{center}
\begin{tabular}{ScScScSrScScSc}
\hline
{\normalsize Case} & \multicolumn{2}{c}{{\normalsize Limits}} &
\multicolumn{4}{c}{{\normalsize Solution Characteristics -- Hall Limit }} \\
& \ \ \ $s_0= \beteo \betio$ & \ \ \ $\Lambda_0=\Upsilon_0 |\betio|$ & 
$|db_r/db_\phi|_0$ &  $\tilde
h$ & $\tilde z_{\rm b}/\tilde h$ &  
$(\j \bdot {\bmath e}^\prime)_0 $ \\ 
\hline 
(i) &
$> 1$ & $> 1/2$ & \ \ \ \ \  $2 \Upsilon_0 \ \qquad (> 1)$ & \ \ \ $a_0/ \epsilon
\Upsilon_0$ &  \ \ \ $(\epsilon  \Upsilon_0)^2/3\sqrt{2}$  &
\ \ \ $\epsilon^2\Upsilon_0/a_0^2$ \\

(ii) & $ > 1$ & $< 1/2$  & \ \ \ \ \ $1/\betio\ \ \quad (> 1)$ &
\ \ \ $2 a_0 \betio / \epsilon $ & \ \ \ $ \epsilon^2/ 6 \sqrt{2} \Upsilon_0
\betio^3$ & \ \ \ $\epsilon^2/ 4 \Upsilon_0 \betio^2 a_0^2$ \\ 

(iii) & $< 1$  & $>1/2$ & \ \ \ \ \ $2 \Upsilon_0 \beteo \betio\ \ (> 1)$ &
\ \ \ $a_0/ \epsilon \Upsilon_0 \betio \beteo $ & \ \ \ $(\epsilon \Upsilon_0 \beteo
\betio)^2/3\sqrt{2}$ & \ \ \ $\epsilon^2 \Upsilon_0 \beteo \betio /a_0^2$ \\

(iv) & $< 1$ & $< 1/2$  &  \ \ \ \ \ $\beteo \qquad (> 1)$  &
\ \ \ $2 a_0/ \epsilon \beteo $ & \ \ \ $(\epsilon \beteo)^2/ 6 \sqrt{2}\Upsilon_0 \betio$ &
\ \ \ $\epsilon^2 \beteo /4 \Upsilon_0 \betio a_0^2$ \\ 
\hline
\end{tabular}
\end{center}
\end{table*}

The ratio $|db_r/db_\phi|_0$ is given by
\begin{equation}
\left| \frac{db_r}{db_{\phi}} \right |_0 = \frac{2 \upo +
\beta}{1 + q\beta^2}
\label{eq:B_rbB_phib}
\end{equation}
(equation~I.115), where $\beta \equiv 1/\betio$. This expression can be
used to approximate $|b_{r{\rm b}}/b_{\phi{\rm b}}|$, the ratio of the
corresponding magnetic field components at the base of the wind. Another
relationship between the field components at $\zt_{\rm b}$ is provided
by
\begin{equation}
b_{r{\rm b}}^2 + b_{\phi{\rm b}}^2 \approx \frac{2}{a_0^2}
\label{eq:rphi_B}
\end{equation}
(equation~I.90). Since $|db_\phi/db_r|_0$ is always $< 1$ in the Hall
regime, one can approximate, to leading order, $b_{r{\rm b}} \approx
\sqrt{2}/a_0$ and $b_{\phi{\rm b}} \approx -\sqrt{2}(1+q\beta^2)/[(2\upo
+ \beta)a_0]$. The assumption $|db_\phi/db_r|_0\ll 1$ also leads to the
following simplified expression for the magnetically reduced scaleheight:
\begin{equation}
\tilde h \approx \frac{2  a_0}{\epsilon}\ \frac{1 + q
\beta^2}{2\upo + \beta}
 \label{eq:ht}
\end{equation}
(equation~I.119). The height of the base of the wind is, in turn, given
by
\begin{equation}
\zt_{\rm b} \approx \frac{a_0 \epsilon}{3 \sqrt{2} \upo}\  \frac{\upo^2
+ (5/2) \upo \beta + \beta^2}{1 + q \beta^2} 
 \label{eq:zb}
 \end{equation}
(see equation~I.120). Finally, the Joule dissipation term in
Table~\ref{table:constraints1} is evaluated from
\begin{equation}
(\j \bdot {\bmath e}^\prime)_0 = 
\frac{\epsilon^2}{4\upo a_0^2}(1+q\beta^2)\left[ 1 + \left (
\frac{2\upo+\beta}{1+q\beta^2}\right )^2 \right ]
\label{eq:joule}
\end{equation}
(see equation~I.122).  The rate of Joule heating at the disc mid-plane
should not exceed the rate of release of gravitational potential energy
at that location. This requirement is expressed by the rightmost
inequality of Table~\ref{table:constraints}, which involves the ratio of
the tidal scaleheight to the disc radius, $h_{\rm T}/r = c_{\rm
s}/v_{\rm K}$. This parameter does not appear explicitly in the
normalized equations for the disc structure (Section~\ref{subsec:gov}),
but it is used in matching the disc solution to a self-similar wind
solution (via equations~\ref{eq:kappa} and~\ref{eq:ele}). We have
verified that this constraint is satisfied by the matched disc/wind
solutions presented in Section~\ref{subsec:global}.

\bsp
\label{lastpage}
\end{document}